\documentclass[]{arXiv} 
\usepackage{draftwatermark}
\usepackage{graphicx}
\usepackage[]{natbib}
\usepackage{txfonts}
\usepackage{booktabs}

\def\lsim{\mathrel{\rlap{\lower4pt\hbox{\hskip1pt$\sim$}}
    \raise1pt\hbox{$<$}}}                
\def\gsim{\mathrel{\rlap{\lower4pt\hbox{\hskip1pt$\sim$}}
    \raise1pt\hbox{$>$}}}                
\begin{document}
\SetWatermarkText{Accepted in A\&A}
\SetWatermarkLightness{0.75}
\SetWatermarkScale{0.75}
\titlerunning{ }
\authorrunning{ }
%
  \title{Dome C UltraCarbonaceous Antarctic MicroMeteorites}
  \subtitle{Infrared and Raman fingerprints}

   \author{E. Dartois 
   \inst{1}\fnmsep\thanks{send offprint requests to emmanuel.dartois@u-psud.fr}
   \thanks{Part of the equipment used in this work has been financed by the French INSU-CNRS program
``Physique et Chimie du Milieu Interstellaire'' (PCMI).}
        \and
        C. Engrand \inst{2}
        \and
        J. Duprat \inst{2}
        \and
        M. Godard \inst{2}
        \and
        E. Charon \inst{2}
        \and
        L. Delauche \inst{2}
        \and
        C. Sandt \inst{3}
        \and
        F. Borondics \inst{3}
          }
   \institute{
   Institut d'Astrophysique Spatiale (IAS), CNRS, Univ. Paris Sud, Universit\'e Paris-Saclay, F-91405 Orsay, France
\and
   Centre de Sciences Nucl\'eaires et de Sciences de la Mati\`ere (CSNSM), CNRS/IN2P3, Univ. Paris Sud, Universit\'e Paris-Saclay, F-91405 Orsay, France
\and
Synchrotron SOLEIL, L'Orme des Merisiers, BP48 Saint Aubin, 91192 Gif-sur-Yvette Cedex, France
   }

   \date{Received January 10, 2017}

 
  \abstract
    {UltraCarbonaceous Antarctic MicroMeteorites (UCAMMs) represent a small fraction of interplanetary dust particles reaching the Earth's surface and contain large amounts of an organic component not found elsewhere. They are most probably sampling a contribution from the outer regions of the solar system to the local interplanetary dust particle (IDP) flux.}
   {We characterize UCAMMs composition focusing on the organic matter, and  compare the results to the insoluble organic matter (IOM) from primitive meteorites, IDPs, and the Earth.}
   {We acquired synchrotron infrared microspectroscopy ($\mu$FTIR) and $\mu$Raman spectra of eight UCAMMs from the Concordia/CSNSM collection, as well as N/C atomic ratios determined with an electron microprobe.}
   {The spectra are dominated by an organic component with a low aliphatic CH versus aromatic C=C ratio, and a higher nitrogen fraction and lower oxygen fraction compared to carbonaceous chondrites and IDPs. The UCAMMs carbonyl absorption band is in agreement with a ketone or aldehyde functional group. Some of the IR and Raman spectra show a C$\equiv$N band corresponding to a nitrile. The absorption band profile from 1400 to 1100 cm$^{-1}$ is compatible with the presence of C-N bondings in the carbonaceous network, and is spectrally different from that reported in meteorite IOM. We confirm that the silicate-to-carbon content in UCAMMs is well below that reported in IDPs and meteorites.
   Together with the high nitrogen abundance relative to carbon building the organic matter matrix, the most likely scenario for the formation of UCAMMs occurs via physicochemical mechanisms taking place in a cold nitrogen rich environment, like the surface of icy parent bodies in the outer solar system. The composition of UCAMMs  provides an additional hint of
  the presence of a heliocentric positive gradient in the C/Si and N/C abundance ratios in the solar system protoplanetary disc evolution.}
{}
%
%

   \maketitle
%

\section{Introduction}
\label{intro}
Micrometeorites represent the largest mass flux of extraterrestrial material falling on Earth \citep[e.g.][]{Love1993, Taylor1996, Duprat2006, Zolensky2006, Prasad2013, Engrand2017}. They sample the interplanetary dust present in the inner solar system, 
 a large fraction coming from Jupiter-family comets together with a contribution attributed to asteroids, and Kuiper Belt and Oort Cloud comets \citep[e.g.][]{Janches2006, Nesvorny2010, Nesvorny2011, Poppe2011, Poppe2016}. The search for and collection of micrometeorites  performed on the Antarctic continent
has allowed  the recovery of particles preserved from significant atmospheric entry heating and which still hold  information from interplanetary dust. This information  otherwise could only be obtained via  sampling by space probes.
Several countries are involved in these collecting expeditions \citep[e.g.][]{Nakamura2005Mineralogy, Duprat2007, Taylor2008, vanGinneken2012, Yabuta2012, Yabuta2017Formation}. 
The Concordia/CSNSM collection contains thousands of Antarctic micrometeorites collected near the French-Italian Concordia station on Dome C during several campaigns performed since 1999 with the help and support of the French Polar Institute Paul-\'Emile Victor (IPEV). Details of the collecting methods can be found in \cite{Duprat2007}.
 A particularly interesting subset in the  collections is the  fraction of  micrometeorites arising from the outer parts of the solar system, known as Kuiper Belt and/or Oort Cloud comet contributions, which are largely undersampled by space missions. These dust particles contain the record of past formation and evolution in the outer solar system. UltraCarbonaceous Antarctic MicroMeteorites \citep[UCAMMs,][]{Duprat2010} and   a subset of chondritic porous interplanetary dust particles (IDPs) are most probably of this class \citep[e.g.][]{Dobrica2010}.

The UCAMMs are exceptionally organic-rich, in particular containing   organic matter that is significantly larger than a micron in size \citep{Dartois2013, Bardin2014, Engrand2015, Charon2017} that is not found in other types of extraterrestrial matter.
The organic content of UCAMMs is high enough to avoid the requirement of prior chemical treatments to be studied in the infrared, and can be compared with the insoluble organic matter (IOM) extracted from meteorites for an insight into their similarities and differences. 
In this work, we report a $\mu$FTIR and $\mu$Raman study of UCAMM samples, and discuss the implications for their origin and significance.
%
%
\begin{figure*}[htbp]
\begin{center}
\includegraphics[width=\columnwidth*2,angle=0]{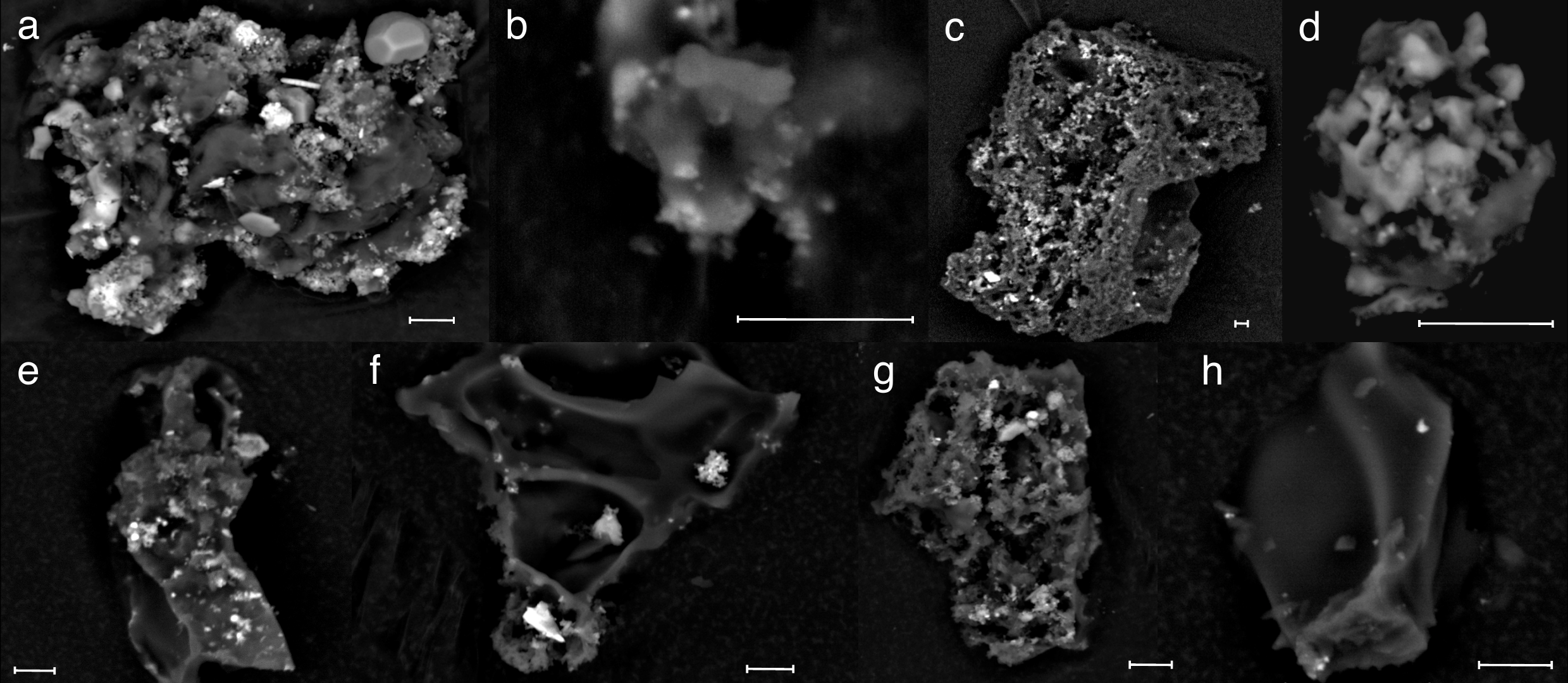}
\caption{Backscattered electron images recorded at 15kV of fragments of the UCAMMs samples analysed in this study: (a) DC060919, (b) DC060594, (c) DC021119, (d) DC060443, (e) DC060718, (f) DC060741, (g) DC0609119, (h) DC060565. Scale bars correspond to 5$\mu$m for each image. The unfragmented UCAMMS initial sizes are respectively: (a) 79 x 50 $\mu$m, (b) 66 x 53 $\mu$m, (c) 110 x 87 $\mu$m (unfragmented), (d) 28 x 24 $\mu$m, (e) 87 x 53 $\mu$m, (f) 200 x 80 $\mu$m, (g) 275 x 108 $\mu$m, (h) 44 x 33 $\mu$m.}
\label{Images}
\end{center}
\end{figure*}
%
%
\begin{figure*}[htbp]
\begin{center}
\includegraphics[width=\columnwidth,angle=0]{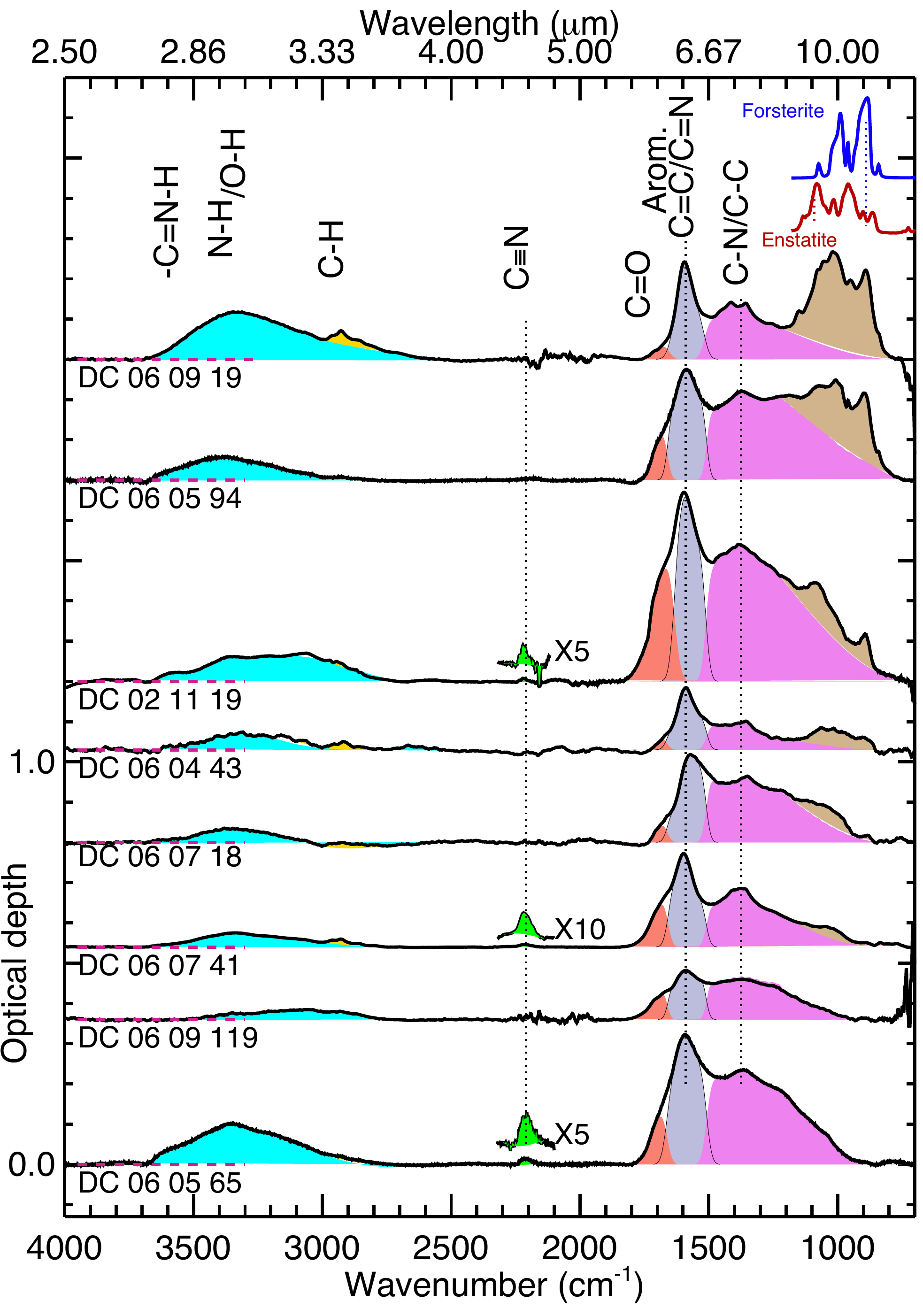}
\includegraphics[width=\columnwidth,angle=0]{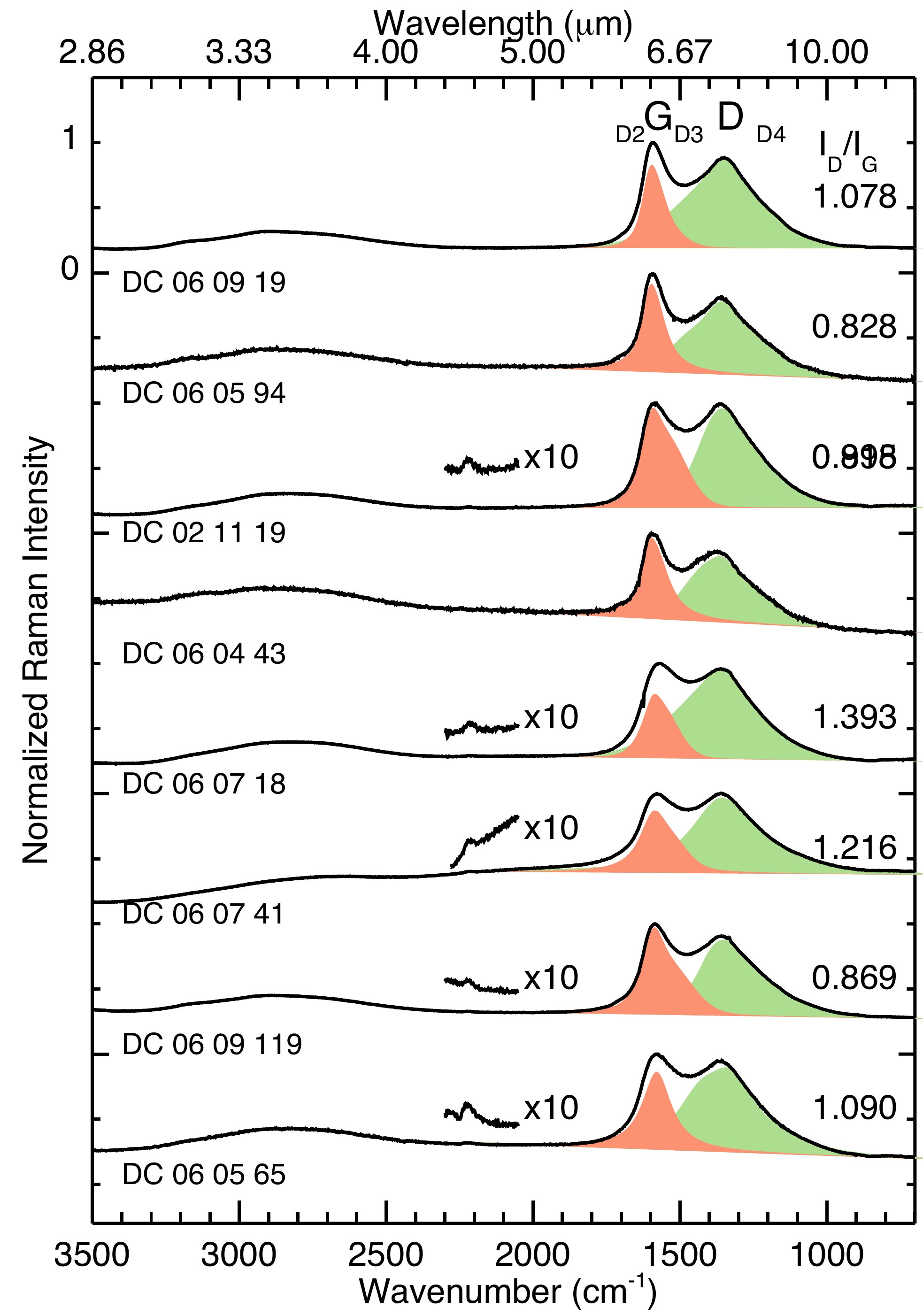}
\caption{Left: UCAMM samples $\mu$FTIR optical depth spectra. The spectra have been vertically shifted for clarity (left horizontal dashed lines indicate the shift level). They are ordered from top to bottom by decreasing amount of silicate band absorption. The main band contributions are labelled above the upper spectrum. The spectra are deconvolved into several contributions shown with distinct  colours, used in the analysis. A zoom on the nitrile region ($\sim$2200~cm$^{-1}$) is provided when a band was measured.
Right: UCAMM samples Raman spectra, normalized to the G-band maximum. The spectra are analysed using a classical Raman bands fitting procedure \citep{Sadezky2005, Kouketsu2014} contributing to the D (green) and G (red) bands. The D/G band peak intensity ratio is given on the right, as defined in \cite{Busemann2007}. A zoom on the nitrile region ($\sim$2200~cm$^{-1}$) is provided when a band was observed. Spectra are vertically shifted for clarity.}
\label{IR_Raman}
\end{center}
\end{figure*}
%
%
\section{Experiments}
%
\begin{figure*}[htbp]
\begin{center}
\includegraphics[width=\columnwidth*2/3,angle=0]{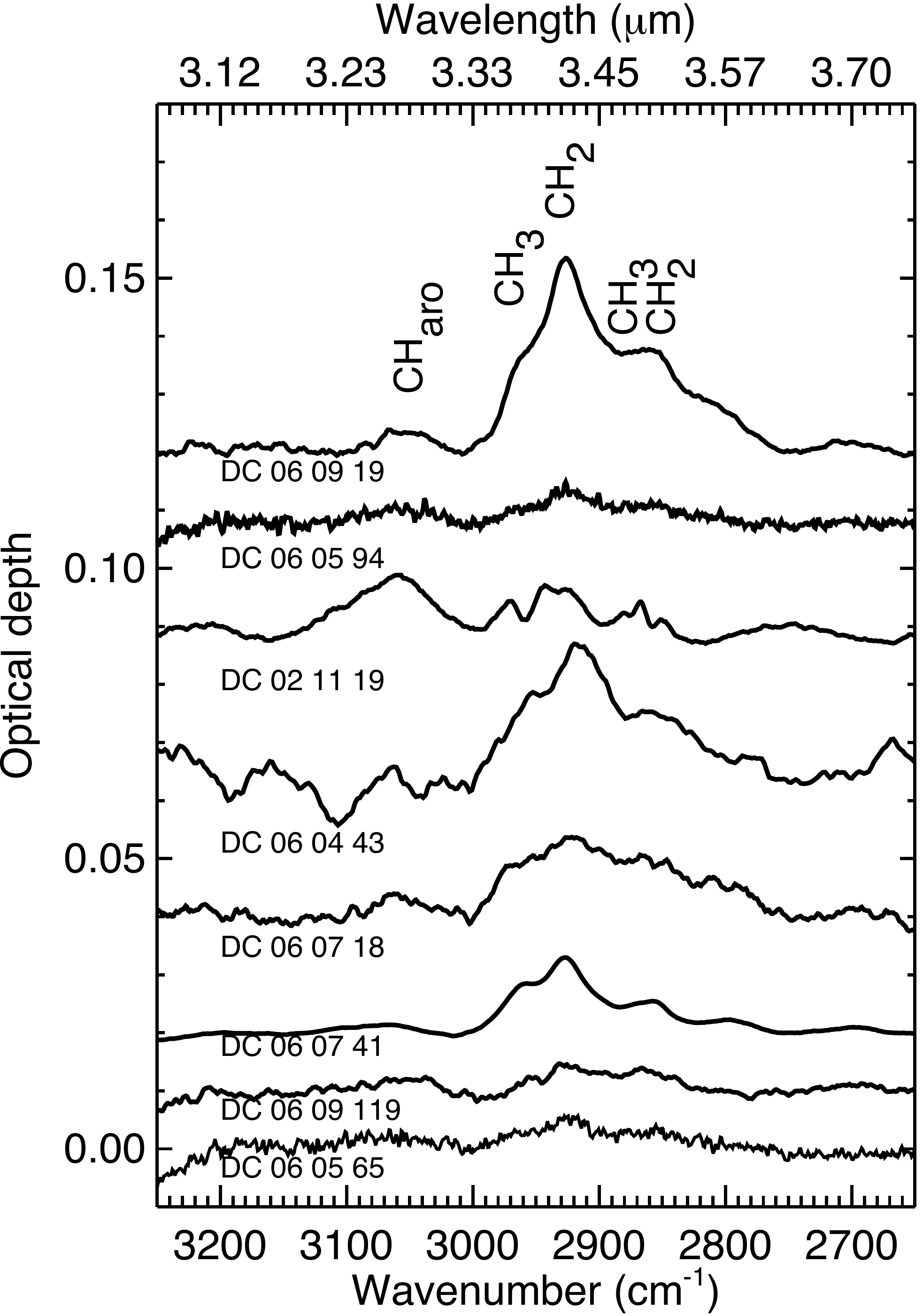}
\includegraphics[width=\columnwidth*2/3,angle=0]{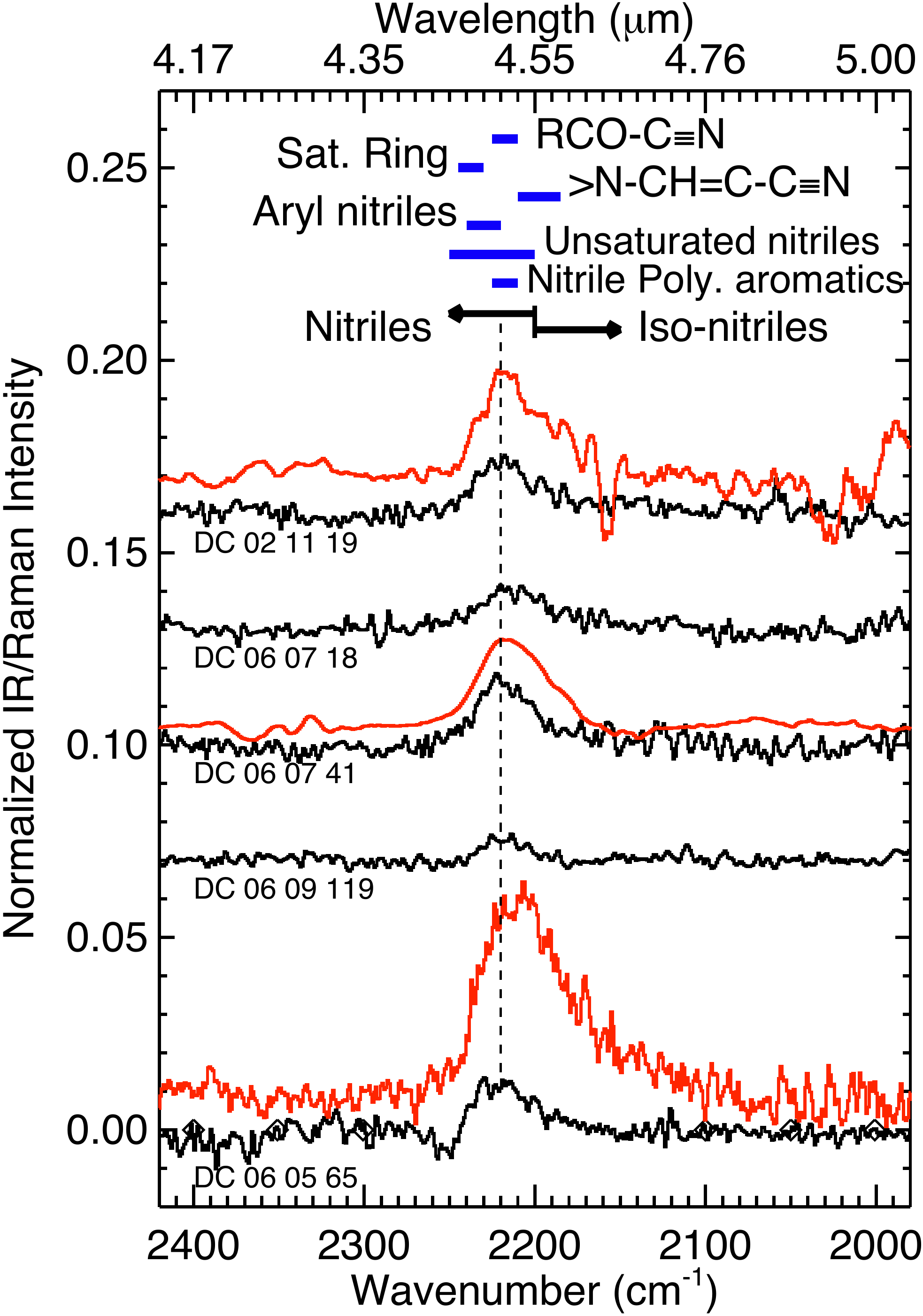}
\includegraphics[width=\columnwidth*2/3,angle=0]{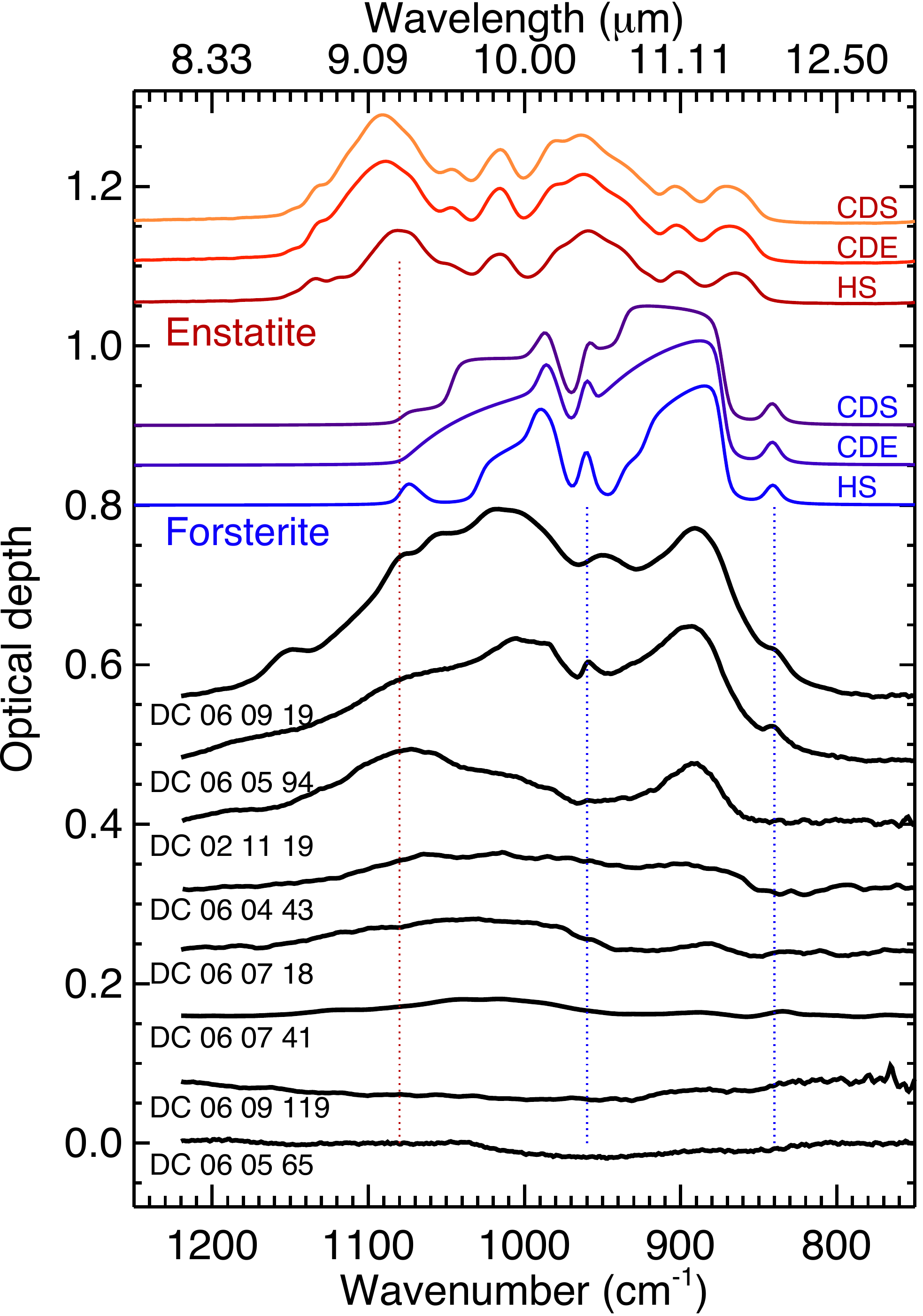}
\caption{Measured UCAMMs spectra. {\it Left}: CH stretching mode infrared optical depth. 
{\it Centre panel}: $\mu$FTIR (red) and Raman (black) close-up spectra in the nitriles region, normalized to the C=C (IR) and G band (Raman) measured intensities.
{\it Right}: Baseline corrected silicate infrared optical depth, along with different absorption models for several size distributions of olivine (forsterite) and pyroxene (enstatite) magnesium rich end members (see text for details).}
\label{CH_Raman_silicates}
\end{center}
\end{figure*}

%
We analysed eight UCAMMs from the Concordia/CSNSM collection.
Fragments of each UCAMM was micro-manipulated and transferred into a diamond compression cell. The fragments were then flattened to provide an optimal thickness for infrared spectroscopic transmission measurements.
The infrared measurements were performed on the SMIS beam line at the synchrotron SOLEIL during several runs in the period from 2010 to 2016. The synchrotron beam was coupled to a Nicolet NicPlan infrared microscope. For the measurements presented in this analysis, the infrared spot size was optimized close to the diffraction limit, with 5x5$\mu$m to 15x15$\mu$m sampling windows, adapted to the geometry of each UCAMM fragment.
A Thermo Fisher DXR spectrometer was used to record a Raman spectrum for each of the samples, using a laser source at 532 nm at minimum power (100~$\mu$W on sample) to prevent sample alteration.
Ten scans with integration times of 20s each were co-added for each spectrum. The spot size of the Raman spectra is lower than that of  the IR spectra, of the order of a micron.\\
For five of the UCAMMS, light element analysis (C, N, O) was conducted at Universit\'e  Paris 6 Jussieu using a Cameca SX100 or SXFive electron microprobe at 10 keV and 40 nA or 20 nA. Before analysis, the samples and standards were coated with a 9 nm  layer of gold using a sputter-coating technique equipped with a quartz thickness monitoring. Care was taken to gold-coat the sample and the standard mounts at the same time whenever possible.
The C, N, and O K$\alpha$ emission lines were recorded using a PC1 crystal (W-Si multilayer crystal), and were calibrated using graphite for C, BN for N, and Fe$_2$O$_3$ for O. As UCAMMs contain variable proportions of minerals intimately mixed with the organic matter; the Na, Mg, Al, Si, P, S, K, Ca, Ti, Cr, Mn, and Fe K$\alpha$ lines were recorded as well, using adequate analysing crystals (LTAP, LPET, or LLIF) and calibrated on usual mineral standards (albite for Na; Olivine for Si and Mg; orthoclase for Al and K; apatite for P and Ca; pyrophanite (MnTiO3) for Mn and Ti; chromite (Cr$_2$O$_3$) for Cr; Fe$_2$O$_3$ for Fe).
Totals different from 100\% can be due to porosity and matrix effects slightly differing between the standard and crushed samples or to a high concentration of an element not measured in the standard (e.g. Ni).
Since these analyses were performed on crushed grains, only analyses with reasonable totals in wt\% (85\% <  totals < 115\%) were selected to calculate the N/ C ratios.
We note that the ratios between C, N, and O measured in a given sample are robust even in the cases of very anomalous totals as they are measured on the same detector.
The typical statistical uncertainties for the determination of C, N, and O in the UCAMMs are on average 4 wt\%, 1 wt\%, and 0.5 wt\%, respectively. The external uncertainty (reproducibility) on the C and N standards are on average 4 wt\% and 2.5 wt\%. The intrinsic variability of a UCAMM sample N/C ratio is usually larger than the analytical uncertainty. 
\section{Analysis}
\label{results}
The measured UCAMM infrared  and Raman spectra are displayed in Fig.\ref{IR_Raman}. The IR spectra are baseline corrected, in a similar way to previously reported in \cite{Dartois2013}, and decomposed into individual contributions corresponding to specific vibrational modes. We also indicate the global fingerprint absorption band in the 1500-800 cm$^{-1}$ spectral region, together with the silicates stretching complex contribution. We represent by different colours the various contributions to the spectra in Fig. \ref{IR_Raman}. 
Except for the $\sim$1100-900~cm$^{-1}$ spectral region, where silicate absorptions prevail if present, the spectra are dominated by the organic matter contribution.
The broad band extending from about 3600 to 3000~cm$^{-1}$ includes contributions from OH and NH stretching modes. The 
asymmetric profile of the band in this range, by comparison with laboratory analogues  \citep{Dartois2013}, indicates a major contribution of the NH stretching mode. 
We measured the amount of nitrogen in DC060594 and DC060565 independently \citep[{N/C ratios of 0.05 and 0.12, locally exceeding 0.15;}][]{Dartois2013}. 
The intrinsic intensity of the OH mode is about ten times higher than the NH stretching mode.
It is probable that the broad and structureless feature between 3300 and 3500 cm$^{-1}$ in DC060919 contains an OH component.\\
A few spectra possess a series of CH stretching mode absorptions, shown in yellow in the left panel of Fig. \ref{IR_Raman}, with a close-up shown in Fig. \ref{CH_Raman_silicates}, left panel. The main bands are due to aliphatic CH stretching modes ($\sim$2960~cm$^{-1}$, asymmetric CH$_3$; $\sim$2920~cm$^{-1}$, asymmetric CH$_2$; $\sim$2870-2850~cm$^{-1}$, symmetric CH$_3$ and CH$_2$). A small contribution at $\sim$3050~cm$^{-1}$ from the aromatic CH stretching mode is observed for the first time in UCAMMs. We also observed a potential small contribution of aldehyde CH at low wavenumber (2820~cm$^{-1}$) in DC060919.\\
Three out of the eight UCAMM fragment analyses show an infrared C$\equiv$N absorption feature at 2200 cm$^{-1}$, which have  also been measured with the Raman spectrometer for five UCAMMs (Fig.\ref{CH_Raman_silicates}, centre panel) and previously observed by \cite{Dobrica2011} for DC060919.
The main vibrational bands associated with the observed IR absorption and Raman emission bands are summarized in Table~\ref{Table1}.

Unlike the other fragment, the DC021119 fragment was retrieved from a previous preparation where it was stuck in a carbon-tape for SEM analysis.
 The DC021119 UCAMM infrared spectrum is therefore affected, mainly in the carbonyl spectral region, and the carbonyl absorption for this UCAMM will thus not be considered in the following analysis.\\
The baseline corrected silicate optical depth contribution is shown in the right panel of Fig.\ref{CH_Raman_silicates}, displayed together with several model absorption spectra of olivine and pyroxene magnesium-rich end members (forsterite and enstatite). The models were calculated with the continuous distribution of ellipsoids (CDE), spheroids (CDS), and hollow spheres (HS) to take into account the variabilities expected due to shape effects in scattering and absorption by randomly oriented particles that are small compared to the wavelength \citep{Min2003}. This statistical approach allows us to compare the amplitude of deformation induced by shapes in the spectra recorded and to infer the silicates composition (one pyroxene and two olivine bands are indicated by vertical dotted lines). It shows notably that the pyroxene band complex still
absorbs significantly at higher wavenumbers (above about 1070-1080~cm$^{-1}$) than the olivine band.
UCAMMs contain a mixture of silicates on a  much smaller scale than that probed by classical IR wavelength microscopy. 
DC060919 and DC060594 display typical  forsterite-like absorption bands, whereas DC021119 absorbs significantly above 1070~cm$^{-1}$, indicating an enstatite-like contribution.\\
%
\begin{figure*}[htbp]
\begin{center}
\includegraphics[width=\columnwidth,angle=0]{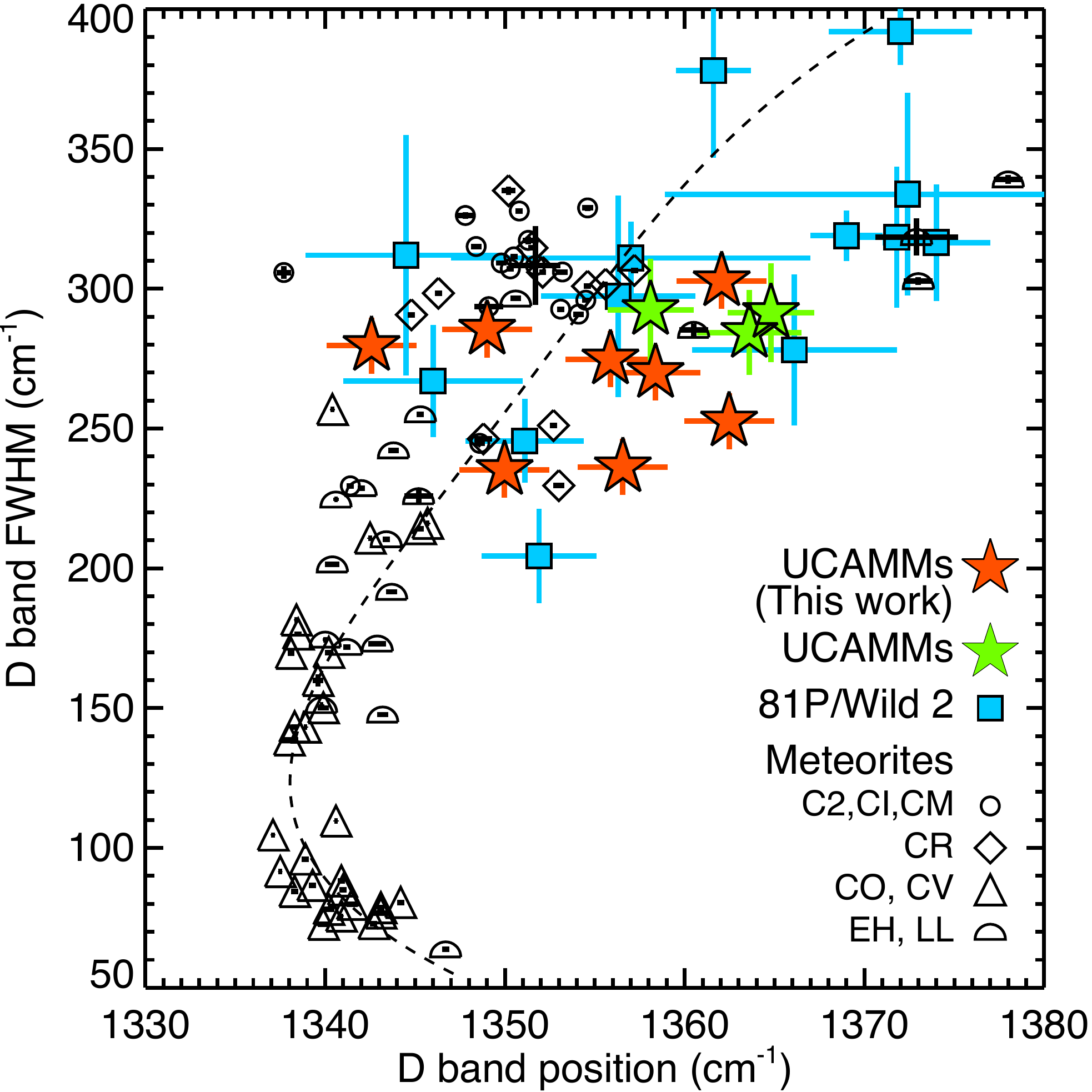}
\includegraphics[width=\columnwidth,angle=0]{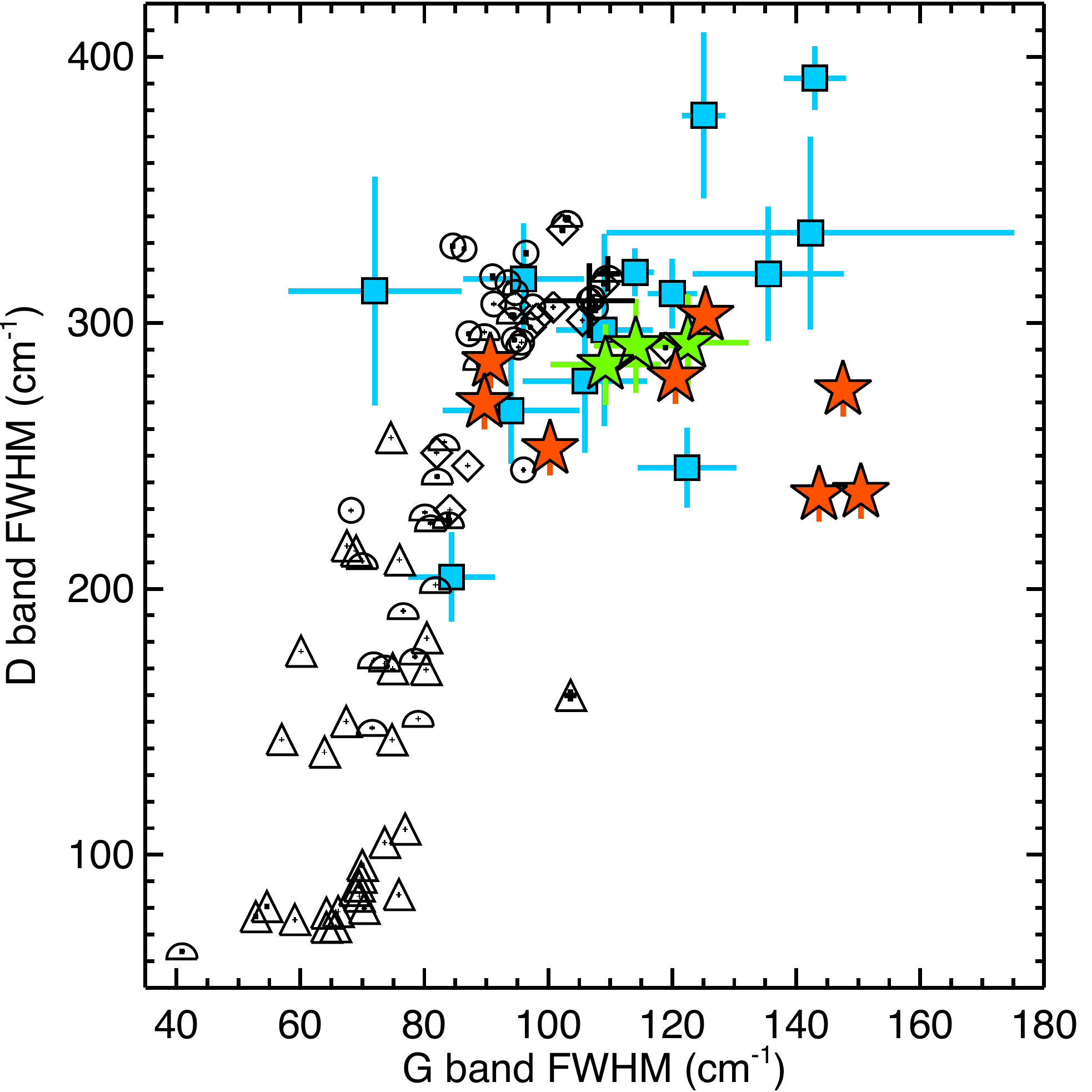}
\caption{Raman D band full width at half maximum (FWHM) versus position ({\it left}) and G band FWHM ({\it right}) diagrams, combining meteoritic IOM \citep[C2,CI,CM: circles, CR: diamonds, triangles: CO,CV \& EH,LL: half circles][]{Busemann2007}, Stardust \citep[squares,][]{Rotundi2008}, previous UCAMM measurements \citep[green stars,][]{Dobrica2011}, and UCAMMs from this work (red stars). The fitted trend is associated with the organic matter primitivity;  the most primitive meteorites have D band width $\gsim$ 250 cm$^{-1}$.}
\label{Raman_D_D}
\end{center}
\end{figure*}
%
To provide a systematic comparison with the organic matter extracted from meteorite IOM, the published infrared spectra from \cite{Kebukawa2011, Orthous2013, Quirico2014} were analysed in the exact same way and are  discussed below.
We also performed infrared measurements  at the synchrotron Soleil SMIS beam line on IOM extracted from the Paris meteorite, kindly provided by V. Vinogradoff  and L.
Remusat (IMPMC/ MNHN)\footnote{The full spectrum will be published by these authors.}.  The analysis of this spectra was performed in the same way and added to the spectra displayed in Fig. 4.
It is difficult to obtain IR spectra covering the full 4000-600 cm$^{-1}$ range for the insoluble organic matter fraction of IDPs. A few exceptions with Hydrofluoric acid (HF) attack of IDPs to eliminate the inorganic fraction, followed by IR measurements, were recorded by e.g. \cite{Matrajt2005}, and are added for comparison to the analysis presented here.\\
The UCAMMs Raman spectra are analysed using a classical Raman bands fitting procedure decomposition contributing to the D and G bands, consisting of the deconvolution of  five sub-bands  \citep[e.g.][]{Sadezky2005, Kouketsu2014} grouped into two main contributions (`D'=D$_1$+D$_2$+D$_4$ and `G'=D$_3$+G), shown in the right panel of Fig. \ref{IR_Raman}. 
\begin{table}
\caption{Infrared and Raman band list}
\begin{center}
\begin{tabular}{|l| l l|}
\hline
Position &\multicolumn{2}{c|}{Mode}\\
(cm$^{-1}$) &IR &Raman\\
\hline
3600-3000       &OH/NH  stretch (br)            &\\
3050            &CH aro.         stretch                        &\\
2960            &CH$_3$ asym. stretch           &\\
2920            &CH$_2$ asym. stretch           &\\
2870            &CH$_3$ sym.     stretch                &\\
2860            &CH$_2$ sym.     stretch                &\\
2220            &C$\equiv$N                             &C$\equiv$N \\
1750-1650       & C=O                                   &\\
1600-1580       & C=C / C=N                             &C sp$^2$\\
1475                    & CH$_{2,3}$ deformation        &\\
1400-1100       & C-N/C-C (br)                          &\\
1375            & CH$_{2,3}$ deformation        &\\
1370-1340       &                                               &C sp$^2$ `Defect' band\\
1100-800                & SiO stretch (Silicates, br)           &\\
\hline
\end{tabular}
\end{center}
\label{Table1}
br: broad
\end{table}%
%
\begin{figure}[htbp]
\begin{center}
\includegraphics[width=\columnwidth,angle=0]{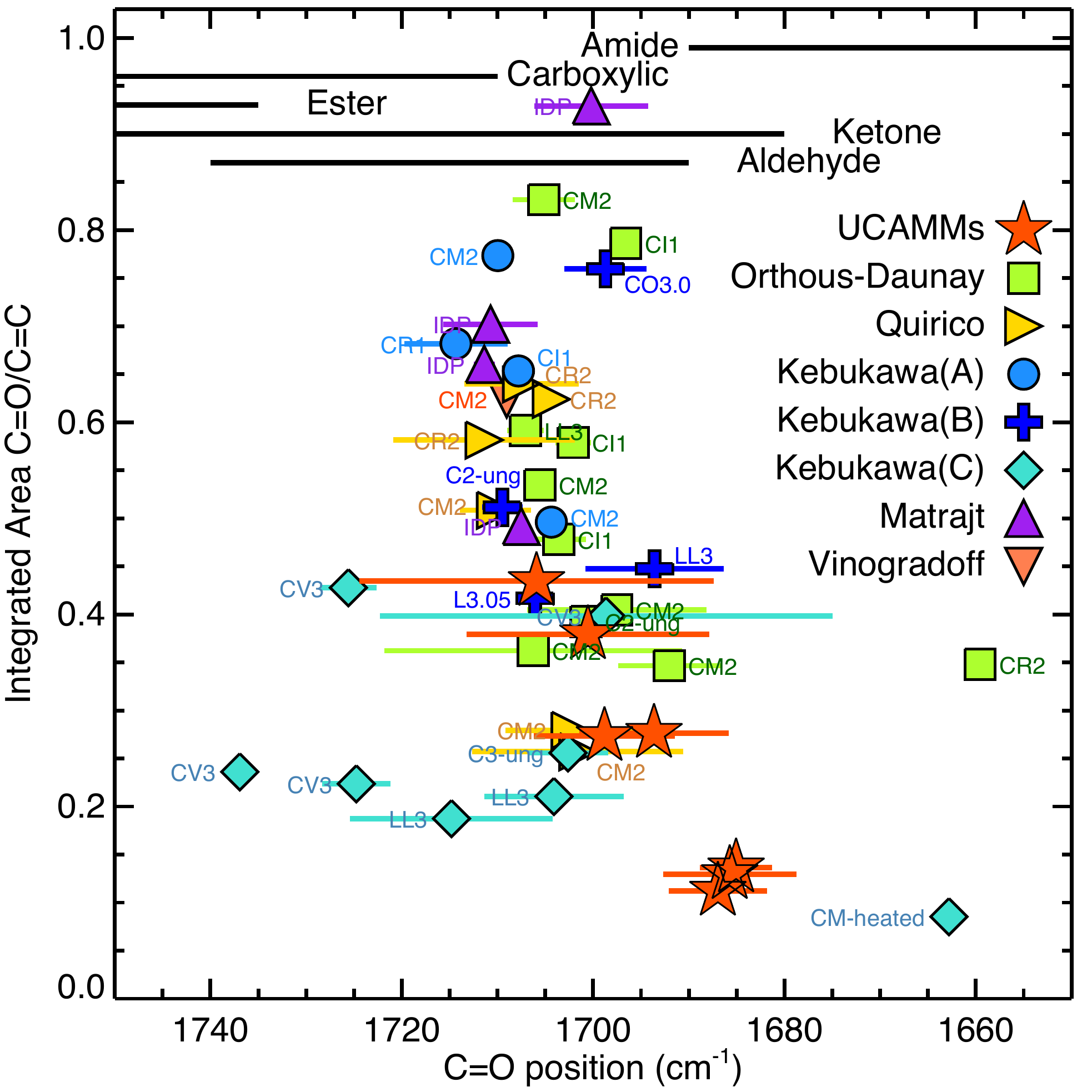}
\caption{Comparison of UCAMM infrared measurements for carbonyl,  displayed with the corresponding infrared features in meteorites.
Carbonyl-to-aromatic-carbon ratio as a function of the position of carbonyl stretching vibrations, and corresponding chemical functional groups.}
\label{IR_combine2}
\end{center}
\end{figure}
%
%
\section{Discussion}

\begin{figure*}[htbp]
\begin{center}
\includegraphics[width=\columnwidth,angle=0]{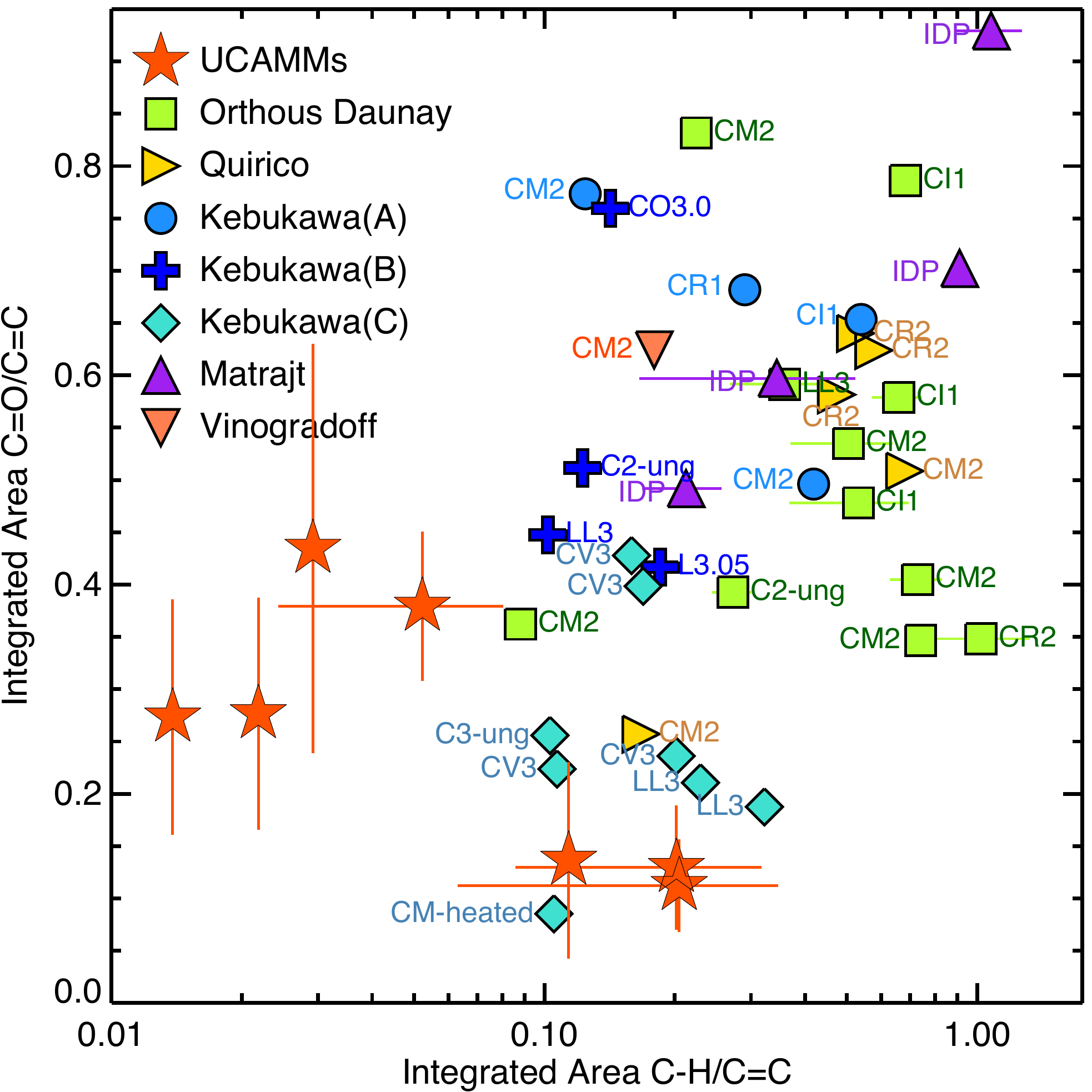}
\includegraphics[width=\columnwidth,angle=0]{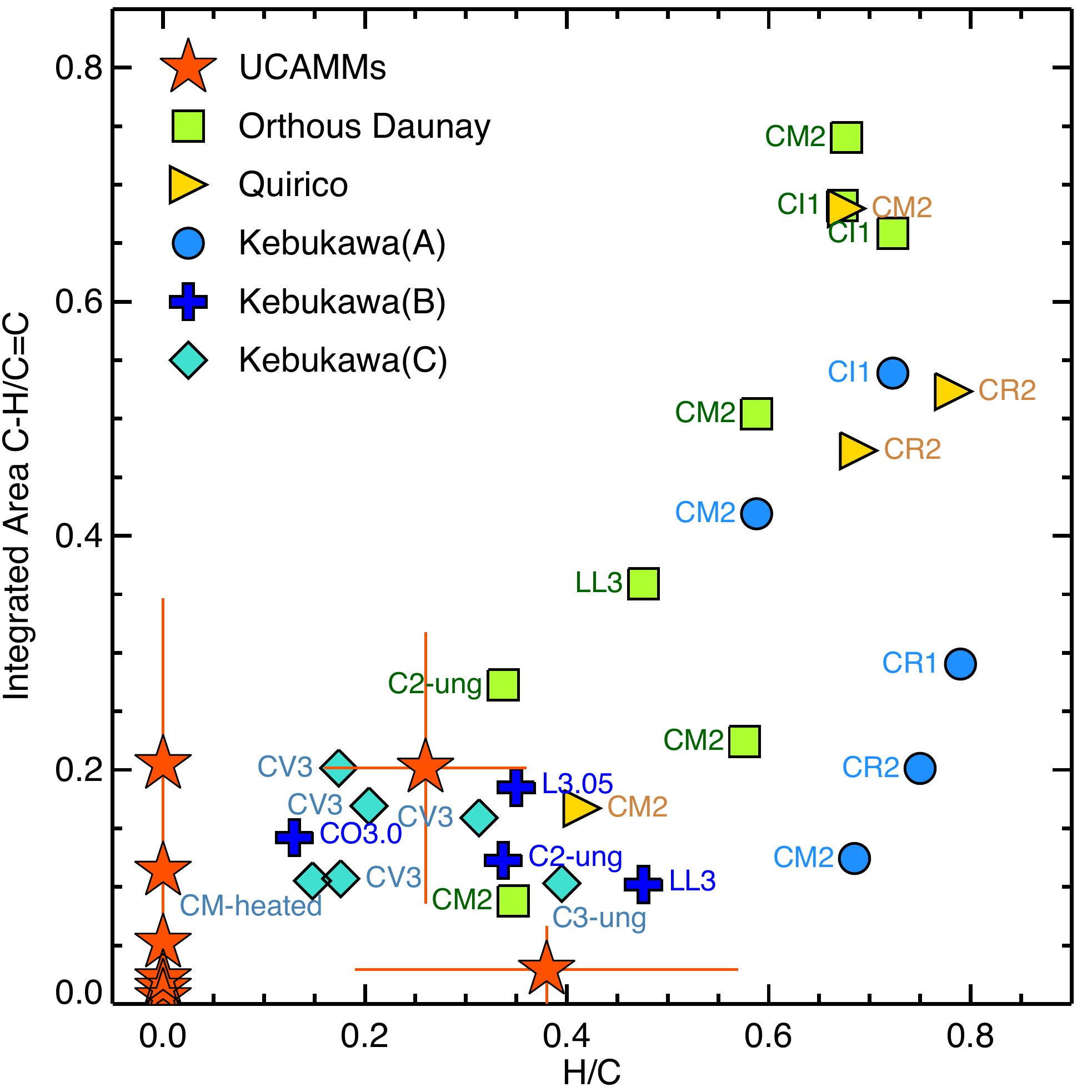}
\caption{Comparison of UCAMM infrared measurements for carbonyl, aromatic carbon, and hydrogen bonded to carbon,  displayed with the corresponding infrared features in meteorite \citep[][separating the  first three infrared subclasses defined in the Kebukawa spectral analysis]{Orthous2013, Quirico2014, Kebukawa2011} and IDP \citep[][]{Matrajt2005} insoluble organic matter. {\it Left}: Carbonyl-to-aromatic-carbon ratio as a function of the CH to C=C areas
{\it Right}: Integrated optical depth area of the CH versus C=C bands as a function of H/C ion probe (NanoSIMS) independent measurements. The UCAMMs without available H/C  measurements are set at a zero H/C value. Labels correspond to meteorite groups, as reported in these articles}
\label{IR_combine}
\end{center}
\end{figure*}

The UCAMMs Raman spectra can be compared to previous measurements for meteoritic IOM \citep{Busemann2007}, Stardust  \citep{Rotundi2008}, and previous UCAMM measurements \citep{Dobrica2011}. The broad and relatively intense Raman D band and intensity ratio of the D and G bands underscore the disordered nature of the carbonaceous network for UCAMMs organic matter. In the D-band full width at half maximum (FWHM) versus D-band position diagram and D-FWHM versus G-FWHM shown in Fig.~\ref{Raman_D_D}, UCAMMs lie in the region associated with the most primitive meteorites and Stardust grains.
\cite{Busemann2007} found and discussed a potential correlation of the atomic N/C element abundance ratios with the D band FWHM \citep[Fig. 8,][]{Busemann2007} and a correlation of the atomic H/C element abundance ratios with the G band FWHM \citep[Fig. 9,][]{Busemann2007}. The UCAMMs measured values are clearly not in line with what would be extrapolated from these correlations, showing that UCAMMs, with their high N/C and low H/C abundances ratios, evolve following another trend.\\
The integrated optical depth of the carbonyl divided by the C=C value is reported in Fig. \ref{IR_combine2} as a function of the carbonyl absorption central position.
A chemical function information arises from the position of the carbonyl. 
Because the electrophile behaviour of neighbouring  atoms has an influence on the bond and thus can alter its position in the spectrum, it is not possible to infer an exact attribution of the carbonyl  by comparison with the expected classical positions. However, as they appear in the 1680-1705 cm$^{-1}$ range, it seems to favour  an aldehyde or ketone carbonyl function over carboxylic acids and ester.
They fall close to the positions of the  primitive classes A and B  from the spectra from \cite{Kebukawa2011},  \cite{Orthous2013}, and \cite{Quirico2014}.\\
The integrated optical depth of the carbonyl absorption band divided by the C=C value is reported in Fig. \ref{IR_combine} (left panel) as a function of the integrated optical depth of the CH stretching mode to C=C ratio. 
We did not include DC021119 in this plot because the tape contamination induces a spectral overlapping contamination in the C=O region. 
The UCAMMs position in this diagram is specific as they clearly lie in the lower left corner with respect to the IOM from meteorites and IDPs.
The low ratio of  C=O to C=C  of the UCAMMs confirms that the O/C ratio in the organic matter of UCAMMs is lower than that of CR and CM meteorites and IDPs.\\
The  low CH to C=C ratio Fig. \ref{IR_combine} (right panel) also indicates a low hydrogen content. In this diagram, UCAMMs for which atomic H/C ratio were not available are set at a zero H/C value.
However, given the general trend observed with the measured infrared CH/C=C band ratio, these UCAMMs most probably are below H/C$\lesssim$0.5.
This value is further confirmed for the two UCAMMs for which a H/C ratio was measured with the NanoSIMS \citep{Duprat2010}. 
The H/C in UCAMMs is substantially lower than those measured for meteorites  from the infrared spectra or independently
by other means by \cite{Alexander2007}.\\
 %
%
The physicochemical composition modifications undergone by meteorites, IDPs, and micrometeorites during their  atmospheric entry 
is a long-standing question \citep[e.g.][]{Greshake1998, Toppani2001, Matrajt2005, Suttle2017}. 
In the Concordia micrometeorite collection, a large fraction ($\sim$35\%) of the particles exhibit textures suggesting that they were not substantially altered during atmospheric entry \citep[e.g.][]{Dobrica2010}. Several indicators point toward a moderate heating for most UCAMMs.
Chemical and structural modifications expected on minerals during thermal processing give a strong upper limit on the temperature suffered by the UCAMMs.
First of all, there is no significant magnetite conversion from pyrrhotite in UCAMMs, estimated to occur in the 500-900$\rm^o$C range \citep[e.g.][]{Rietmeijer2004,Craig1974}. The annealing of tracks in the minerals occurs at a temperature below 600$\rm^o$C. Some of the minerals in UCAMMs display these tracks \citep[][]{Charon2017}.
The high organic content of UCAMMs is also an indicator of at most moderate heating.
The UCAMM organic matter Raman G and D band FWHMs are large;  for some UCAMMs they lie above 140~$\rm cm^{-1}$ for the G band, the largest among extraterrestrial organic matter, which points toward a low atmospheric entry heating by comparison to experimental simulations \citep[e.g.][]{Bonnet2015}.
The high amount of nitrogen in UCAMMs organic matter and the presence of NH stretching modes confirm the low heating;  above about 500$\rm^o$C these NH bonds should be reduced drastically, and this implies that the N/C would otherwise be even higher before entry. The  greatest unknown is probably the initial C-H content in UCAMMs, which may be reduced during  atmospheric entry. 
IDPs often preserve higher aliphatic C-H content in their organic matter than measured for UCAMMs. Because  IDPs are usually smaller in size, they may have suffered less heating during the atmospheric entry \citep[e.g.][]{Love1991}. 
Undoubtedly ices, including nitrogen ices, are  lost long before the atmospheric entry, in the evolving parent body, and during the journey of the dust grains toward the inner solar system. This evolution most probably helped to form their original macromolecular organic content as discussed in the UCAMMs formation scenarios \citep{Dartois2013, Auge2016}. 
The nitrogen budget of UCAMMs before atmospheric entry may be higher than measured in this study if they suffered some heating during atmospheric entry. The nitrogen-to-carbon atomic ratio is indeed expected to decrease with the heating of the particles \citep[e.g.][]{Bonnet2015}, but should remain relatively stable up to about 500$\rm^o$C. Above such temperatures, the polyaromatic network is also strongly modified and the nitrile infrared signature disappears, whereas it is observed in some of the UCAMMs in this study. 
If the particles were heated to much higher temperatures, the measured N/C ratios should be considered as lower limits of the initial N/C ratios as heating would result in nitrogen loss. The chemical element the  most prone to substantial loss from the organic content of UCAMMs is hydrogen as it is the most labile.
The CH content in UCAMMs is variable, as measured in the infrared spectra, and in some cases relatively low.
It is therefore not possible to definitely rule on the pre-atmospheric entry CH content of UCAMMs. It may be higher before atmospheric entry
than measured in this study and reduced upon atmospheric entry.
 It could also be the initial and preserved CH content, with the hydrogen being mainly bound as NH, as currently observed.
However, taken together, the above-mentioned arguments and measurements provide a substantial body of evidence that UCAMMs probably did not suffer a temperature of more than about 100-500$\rm^o$C during atmospheric entry.

\subsection*{ Nitrogen contribution and the N/C solar system gradient}

\begin{figure*}
\begin{center}
\includegraphics[width=\columnwidth,angle=0]{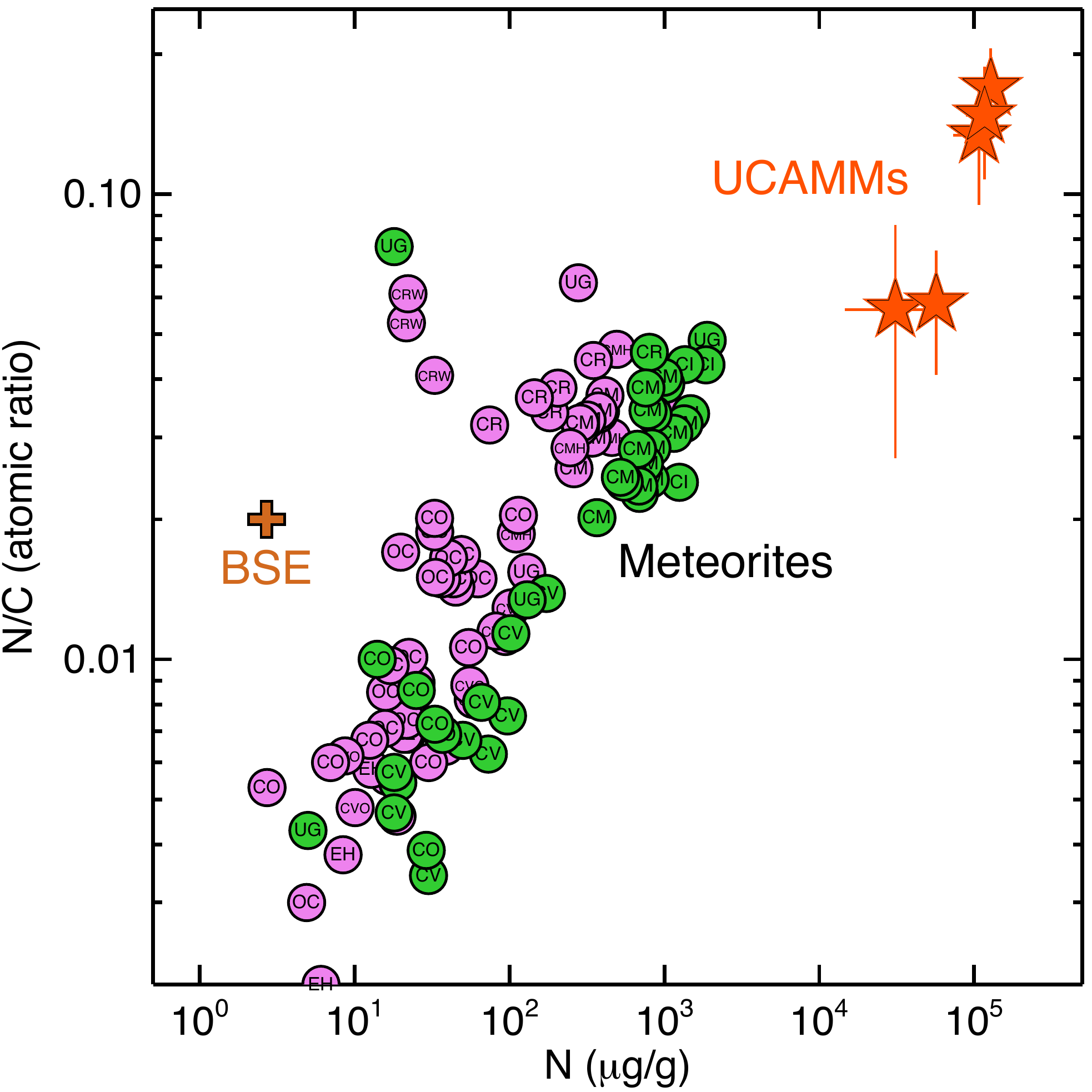}
\includegraphics[width=\columnwidth,angle=0]{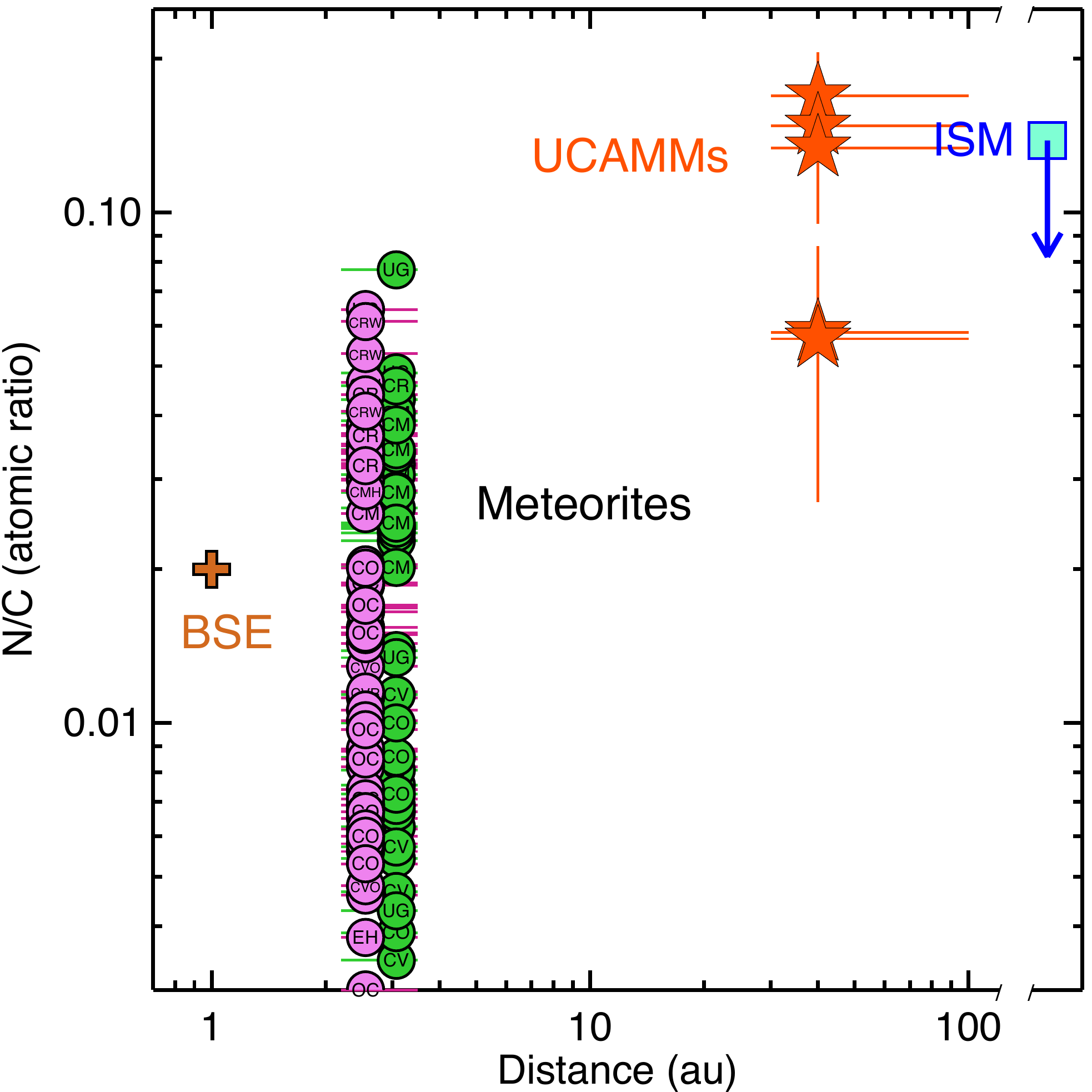}
\caption{UCAMM nitrogen abundances relative to carbon (atomic ratio) are compared to solar system solids: meteorites \citep[][green and purple circles, respectively; labels correspond to meteorites groups as reported in these articles]{Kerridge1985, Alexander2007}, the Bulk Silicate Earth \citep[BSE,][]{Bergin2015}, and the interstellar medium (see text for details). Left: Reported in function of bulk nitrogen weight fraction; Right: Reported in function of heliocentric distance.
Figure \ref{N_C}  summarizes the distance, N/C, and N weight fraction information in a single plot.
UCAMMs values are reported in Table~\ref{Table_summary}.}
\label{N_C_double}
\end{center}
\end{figure*}

The position of the C$\equiv$N band measured  in the IR and in the Raman spectra, shown in the central panel of Fig.~\ref{CH_Raman_silicates}, corresponds to a nitrile function; an isonitrile group would absorb at lower frequencies \citep[e.g.][]{Bonnet2015}.
The absorption profile in the fingerprint region (1500-800 cm$^{-1}$) is significantly different in UCAMMs compared to meteorites' IOM.
It is particularly evident when the organic content is the highest and the silicates represent a negligible fraction of the absorption profile (lower spectra of left panel in Fig.\ref{IR_Raman}). The absorption bands peak around 1380~cm$^{-1}$, whereas in IOM spectra the maximum of the broad absorptions peaks around 1200~cm$^{-1}$  \citep[e.g. Fig.~1][]{Dartois2014, Quirico2014, Orthous2013, Kebukawa2011}, with variable methyl and methylene deformation mode contributions on top of the profile at higher wavenumbers. This UCAMM fingerprint region profile is reminiscent of the nitrogen-rich carbonaceous network, as observed in the spectra of laboratory a-CNH analogues \citep[e.g.][]{Auge2016, Quirico2008, Lazar2008, Gerakines2004, Fanchini2002, Rodil2001, Hammer2000, Ong1996}, in agreement with the higher nitrogen fraction in the organic phase of UCAMMs \citep[][]{Dobrica2011, Dartois2013}. 
If the main nitrogen  contribution to the UCAMMs is not the C$\equiv$N, but C=N and C-N bonds contributing to the infrared spectra at lower frequencies, what makes the C$\equiv$N feature unique is that it falls in a clean spectral region, and when measured in organic matter, it is in most cases the sign of a high nitrogen content.
UCAMMs are probably one of the poles in the nitrogen history of the solar system solids, probably sharing commonalities with some of the nitrogen-rich hot spots found at much smaller scales (i.e. lower organic matter fraction) in some IDPs \citep[e.g.][]{Aleon2010}.
The nitrogen-to-carbon abundance ratio in UCAMMs, measured with an electron microprobe for five of them, are shown in Fig. \ref{N_C_double} (left panel) together with their nitrogen absolute abundances (in weight fraction), and are compared to other solar system solids.
The N/C is one of the indicators used to discriminate between different reservoirs for the organic matter in solar system solids.
Although difficult to evaluate, the Bulk Silicate Earth (BSE) N/C value lies about an order of magnitude lower \citep[][and references therein]{Halliday2013, Marty2012, Bergin2015}. 
The interstellar medium N/C is somewhat more difficult to constrain as---in contrast to measurements on interplanetary carbonaceous dust---nitrogen in diffuse ISM solids is spectroscopically elusive. Nevertheless, an upper limit  can be determined assuming that the depletion of nitrogen and carbon observed in the diffuse medium is locked into carbonaceous dust.  
Considering the missing fraction of nitrogen ($\rm \delta N$)  one can form the elemental ratio $\rm N/C \lesssim(\delta N \times [N])/(\delta C \times [C])$ $\rm\approx (0.2\times80ppm)/(0.4\times290ppm)\approx0.14$ \citep[e.g.][and references therein]{Verstraete2011}. 
This ratio represents a stringent upper limit for the N/C ISM value that is in agreement with the fact that spectroscopic observation of both the aromatic infrared bands emission carriers, also called `astrophysical PAHs' (polycyclic aromatic hydrocarbons), and hydrogenated amorphous carbon solids in the ISM do not show a large incorporation of nitrogen as an heteroatom. Abundances of 
N/C  in the organic matter from measurements for meteorites from the asteroid belt \citep[][]{Kerridge1985} are also shown in Fig.\ref{N_C_double}.  Their N/C ratio and absolute N abundance are both lower than in UCAMMs, with a slight overlap for the extremes in each distribution.
The bulk N abundance in meteorites is lower than that of UCAMM by up to an order of magnitude.
UCAMMs are placed in a N/C versus estimated heliocentric distance diagram in the right panel of Fig.\ref{N_C_double}.
As discussed in \cite{Bergin2015, Millar2015, Lee2010}, the N/C ratio is sensitive to the disc chemistry and the radial transport, with a greater retention of C over N for thermal or impact events.
Disc models \citep[e.g.][]{Piso2016} endeavor to demonstrate to what extent the C/O and N/C ratio can be related to the abundance of specific carriers, the composition of the ice (dominated or not by water ice, which controls the binding energy of more volatile ices), and the effect on the temporal evolution of disc temperatures on the position of the snow lines of the  volatile species. 
Up to the planetesimals formation phase, observations show outer regions dense and cold enough to condense even the most volatile species.
This includes the nitrogen molecule that is unfortunately weakly infrared active via ice interaction induced transitions, and thus escaping to date a direct abundance determination in ice mantles in this phase. These condensation phases are indirectly traced by the observation of related species in the gas phase, such as the N$_2$H$^+$ radical.\\
%
%
%
%
\begin{figure*}[htbp]
\begin{center}
\includegraphics[width=\columnwidth,angle=0]{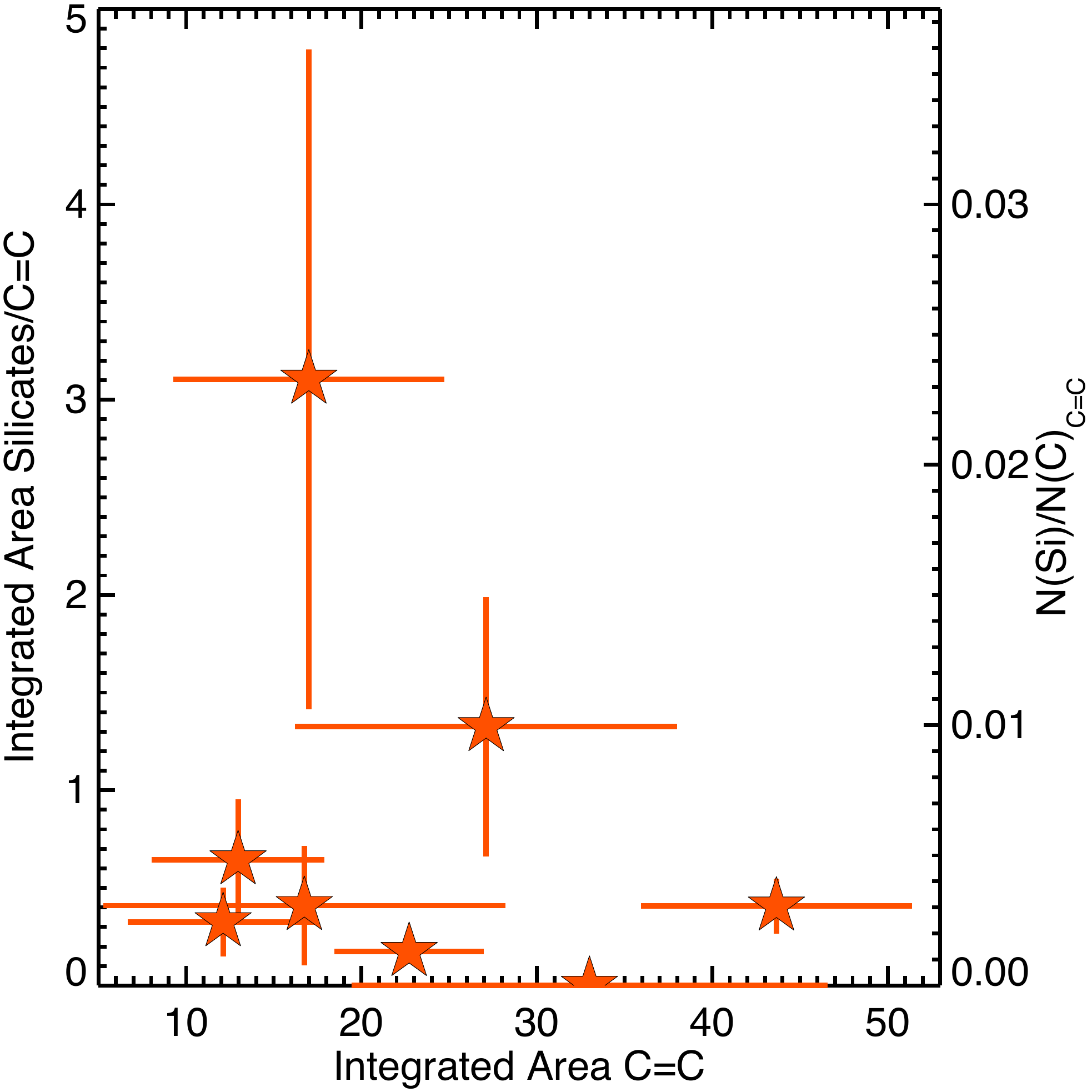}
\includegraphics[width=\columnwidth,angle=0]{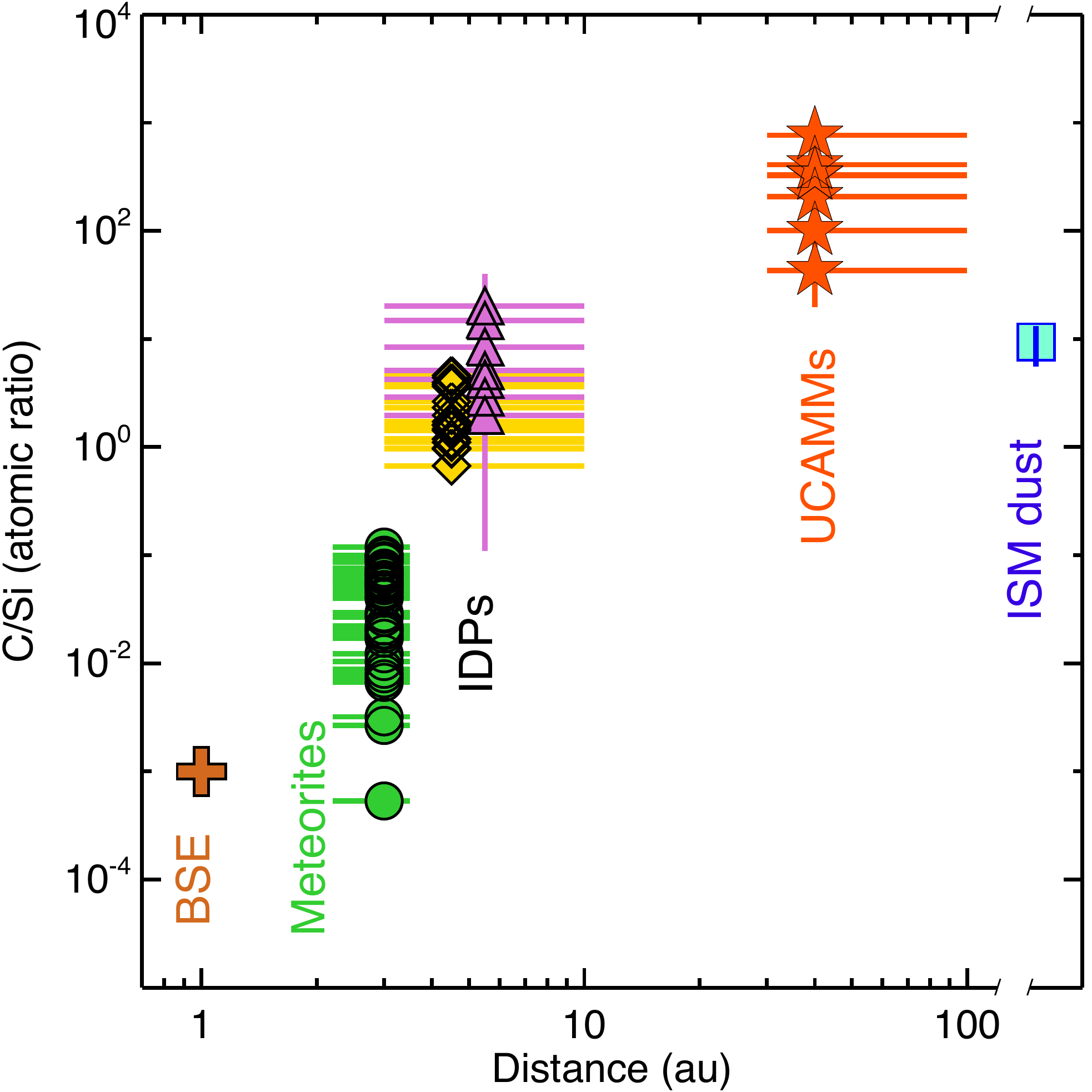}
\caption{{\it Left}: Silicates-to-aromatic C=C integrated area ratio and corresponding silicon-to-carbon atomic ratio (right axis, see text for details) in UCAMMs.  {\it Right}: UCAMMs (stars) carbon abundances relative to silicon (atomic ratio) are compared to solar system solids: IDPs \citep[][triangles and diamonds, respectively]{Matrajt2005, Thomas1993}, meteorites \citep[][circles]{Kerridge1985}, the bulk silicate Earth \citep[BSE,][cross]{Bergin2015}, and the interstellar medium (square, see text for details).}
\label{C_Si}
\end{center}
\end{figure*}
%
%
Following this early phase, and for up to 4 Gyrs, larger icy bodies at large heliocentric distances can retain volatiles such as CH$_4$
 and N$_2$  on their surfaces.  An emblematic object of this type is Pluto, with a detailed spatial record of CH$_4$- and N$_2$-rich ices at its surface, recently mapped by the New Horizons mission \citep[e.g.][]{Protopapa2016}. Different terrains and ice compositions were observed, and the dynamical history of Pluto seems to favour ices segregations, with regions stable for very long periods of time, possibly over the age of Pluto itself \citep[][]{Hamilton2016}. Pluto is among the largest bodies with (sub)surface CH$_4$  and N$_2$ volatile ices; however, in the Kuiper belt and the Oort cloud regions many smaller icy bodies should have (sub)surface N$_2$ and CH$_4$ dominated ices  \citep[e.g.][]{Brown2011}. These surfaces are exposed to Galactic cosmic-ray irradiation, and macromolecular precursors can be synthesized by irradiation of such ices \cite[see the scenarios and experiments detailed in][]{Dartois2013, Auge2016}. 
In UCAMMs we see little evidence of a pristine, directly incorporated and unmodified, interstellar dust organic component. If the UCAMM organic matter was produced closer to the Sun by an alternative scenario, it would be difficult to reconcile with the high N/C and D/H and lower O/C ratios than found in
meteorite IOM. The overall results indicate that the UCAMM bulk organic matter is not a direct heritage from and incorporation of the presolar cloud.
\subsection*{The C/Si solar system gradient}
The silicon-to-carbon abundance ratio in UCAMMs can be evaluated by forming the ratio of the integrated silicates band around 1000~cm$^{-1}$ to the contribution from the C=C band around 1590~cm$^{-1}$, which largely dominates the carbon network.
Using the integrated absorption cross sections of the silicates band of $\rm 1.8\pm0.2\times10^{-16}$~cm/Si \citep[][]{Matrajt2005, Bowey2002, Dartois1998} and that of the C=C band  of $\rm 1.5\pm0.5\times10^{-18}$~cm/C evaluated in \cite{Dartois2013}, this ratio can be converted into Si/C values. They are reported in the right ordinate of Fig. \ref{C_Si}, showing that the UCAMM silicates content is low (Si/C $\lsim$ 0.03). This Si/C ratio measured in the infrared confirms the ultracarbonaceous nature of the fragments analysed.
In this figure are displayed IDPs measurements of the Si/C by \cite{Thomas1993} and values retrieved from the infrared spectra from \cite{Matrajt2005}.
For the latter, the C/Si value reported for each of the seven IDPs analysed, is the mean of the C/Si given in Table 4 of \cite{Matrajt2005}, including only aliphatic carbons, and a reevaluation based on the infrared spectra of the maximum C possibly hidden in the C=C band absorption region. The vertical lines indicate the full range between the two calculated values.
The major flux of such IDPs is attributed to parent bodies coming from Jupiter-family comets and asteroids, with heliocentric distances in the 3-10 AU range \citep[][]{Poppe2011, Poppe2016}. 
Measurements for meteorites from the asteroid belt \citep[][]{Kerridge1985} are also shown.
The bulk silicate Earth \citep[BSE,][and references therein]{Bergin2015} C/Si value lies orders of magnitude below. 
The C/Si can also be evaluated from interstellar medium measurements. An upper limit can be determined by simply forming the cosmic abundance ratio for carbon and silicon, which is around 10, based on recent measurements \citep[][]{Jenkins2014, Jenkins2011, Jenkins2009, Przybilla2013}.
The lower bound can be estimated for carbon by looking either at the depletions, attributed mainly to solids formation, or also by quantifying the spectroscopically identified carbonaceous dust \cite[e.g.][and references therein]{Dartois2015} or the carbon fraction required in dust models \citep[][]{Jones2015, Draine2015, Wang2014}. Most of the silicon is locked into the formation of inorganic compounds (mainly amorphous silicates in the diffuse interstellar medium; \citealt{Kemper2004}). The lower bound is therefore around 40\%\ of the cosmic abundance.
We report the C/Si ratio as a function of the estimated heliocentric distance for the different objects considered (Fig. \ref{C_Si}, right panel), and compare it to the interstellar medium. The UCAMMs display the highest ratios above the meteorites and, importantly also above the interstellar value, which is  considered  the maximum direct C/Si abundance heritage value.
As stated in \cite{Millar2015}, the carbon abundance appears to be closer to normal value (i.e. C/Si cosmic abundance) in the cooler materials that reside in the outer solar system, but globally the incorporation of carbon into refractory bodies is difficult.  The preferential destruction of carbon grains over silicates in the inner protosolar disc \citep[][]{Lee2010} would naturally explain a carbon deficiency gradient.\\
Recent models propose that a large fraction of carbon in the early solar system was removed from the dust component, especially in the inner regions of the solar system. These models thus predict a radial dependence of the abundance of carbon dust and could explain the depletion observed for the carbon abundance in planetesimals in the asteroid belt and in the terrestrial planet region \citep[e.g.][]{Gail2017Spatial}
These variations may have observational consequences for protoplanetary discs. In the remote observations of protoplanetary discs, PAH emission lines and  carbonaceous dust carriers in general, are conspicuous by their absence, whereas silicates are clearly observed in emission in the zones approximately corresponding to  the sizes of the actual inner solar system. In addition to the inherent emissivity lower contrast in bands and/or lines for many carbonaceous solids  compared to silicates, such a C/Si radial variation could play an important role in the difficulty encountered to detect them.\\
Finding a high C/Si value in an extraterrestrial dust grain---as is found in UCAMMs (C/Si $\gsim$ 10) and some IDPs---associated with a high N/C ratio \citep[][]{Duprat2010, Dartois2013}, can thus most probably be assigned to an incorporation via a mechanism occurring in the outer solar nebula, such as the proposed irradiation scenario for UCAMMs; this is suggested by  the C/Si gradient and  also because the direct incorporation from the diffuse ISM would apparently lead to a lower C/Si value.

\section{Conclusion}

Infrared and Raman $\mu$spectroscopy spectra of eight UCAMMs fragments provide a more comprehensive picture on the physicochemical composition of UCAMMs.
The UCAMMs clearly contain a large amount, usually well over a micron in size, of N-rich organic matter that is not found in other types of extraterrestrial matter.
The high carbon abundance relative to silicon in UCAMMs is well above that of most solar system primitive samples (IDPs, meteorites). It is also, for some of them, significantly above the mere interstellar C/Si value which is probably not compatible with the hypothesis of a direct incorporation of an interstellar precursor.
This favours scenarios in which the N enrichment is acquired as a secondary  physicochemical process occurring during the protoplanetary phase or later, and not a heritage from pristine incorporated interstellar matter.
The UCAMMs micrometeorites reveal the existence of nitrogen-rich precursors formed beyond the nitrogen snow line, and a chemistry regime occurring in the solar system in outer regions.
UCAMMs are additional evidence 
for the presence of a positive gradient of the C/Si and N/C abundance ratios with heliocentric distance, as expected
from protoplanetary disc evolution models and progressively revealed by astrophysical disc observations.
 
\begin{acknowledgements}
We acknowledge SOLEIL for the use  of the synchrotron radiation facilities. 
The authors warmly thank Paul Dumas for the discussions and support over many years and for the successful realization of these experiments.
The authors  are grateful to Bruno Crane, Nicolas Szwec, and Silvin Herv\'e for their help in the mechanical designs used to perform the experiments.
We acknowledge Y. Kebukawa for kindly providing us with her original data.
The authors would like to cordially thank Martin David Suttle and Tristan Guillot for their constructive comments and the language editor Helenka Kinnan for suggestions, improving the scientific content quality and readability of this article.
Part of the equipment used in this work has been financed by the French INSU-CNRS program ``Physique et Chimie du Milieu Interstellaire'' (PCMI). 
The measurements of the N/C with the electron microprobe were performed thanks to the CAMPARIS team at Jussieu.
This work was supported by the ANR projects COSMISME (Grant ANR- 2010-BLAN-0502) and OGRESSE (Grant ANR2011-BS56-026-01) of the French Agence Nationale de la Recherche as well as INSU, IN2P3, CNES, DIM-ACAV (R\'egion Ile de France), CNRS, and Universit\'e Paris-Sud. This work is part of the JWST emblematic project from the LABEX-P2IO.
We are also grateful to the French and Italian polar institutes IPEV and PNRA, for their financial and logistic support of the micrometeorites Concordia collection.
\end{acknowledgements}

\newpage
\newpage

\appendix

%
\section{N/C extended version}
%
%
%
\begin{figure}
\begin{center}
\includegraphics[width=\columnwidth,angle=0]{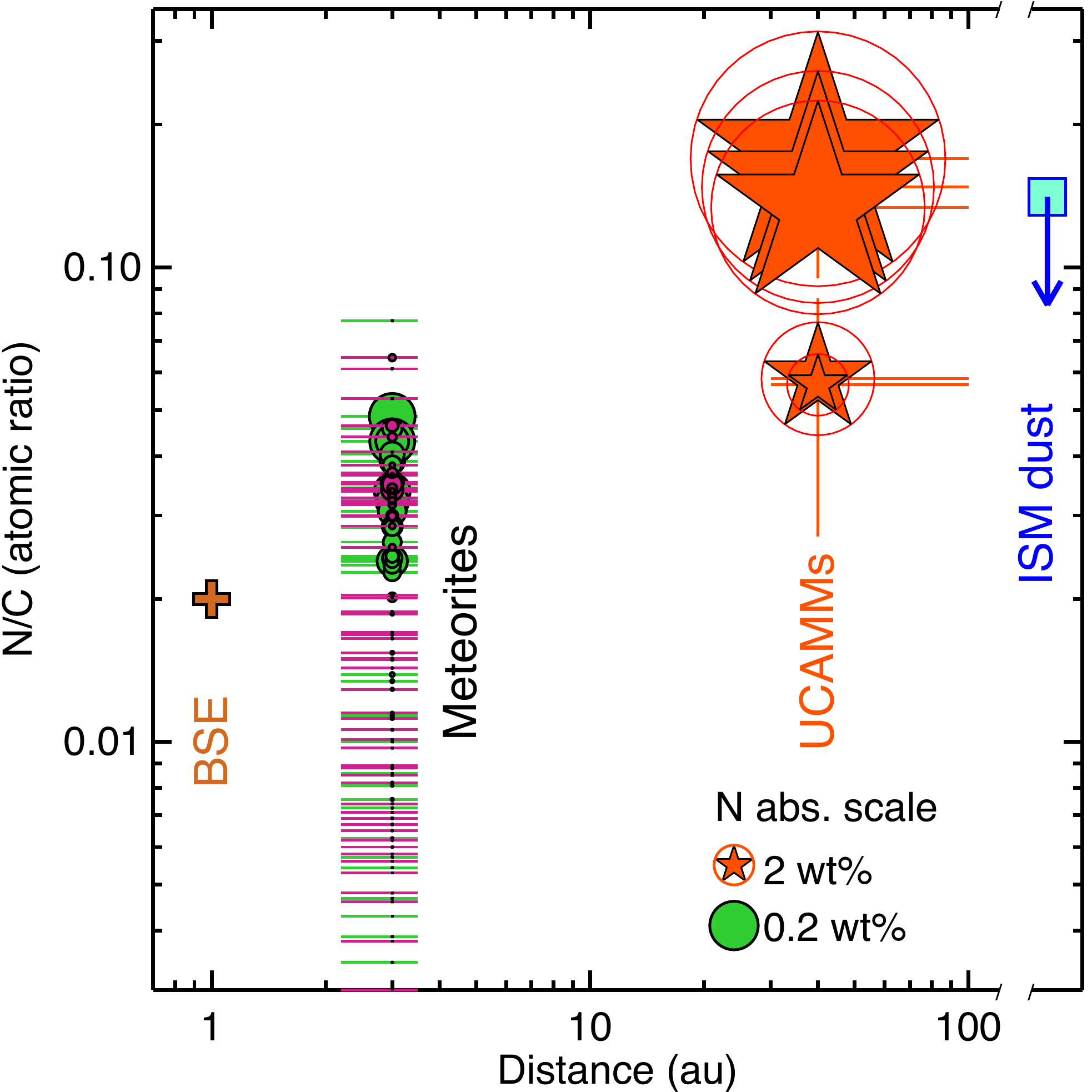}
\caption{UCAMM nitrogen abundances relative to carbon (atomic ratio) are compared to solar system solids: meteorites \citep[][]{Alexander2007, Kerridge1985}, the bulk silicate Earth \citep[BSE,][]{Bergin2015}, and the interstellar medium (see text for details). The sizes of the symbols for UCAMMs and meteorites give the absolute nitrogen content (in wt \%). UCAMM values are reported in Table~\ref{Table_summary}.}
\label{N_C}
\end{center}
\end{figure}
%
\section{Measurements summary}
\ref{Table_summary}
%
%
\begin{sidewaystable*}
\caption{Summary}
\begin{center}
\begin{tabular}{l l l l l l l l l}
\hline
UCAMM                           &DC060919       &DC060594               &DC021119       &DC060443                 &DC060718               &DC060741               &DC0609119      &DC060565 \\
\hline
\multicolumn{9}{l}{Raman parameters (estimated uncertainty $\sim$ position$\pm$2.5cm$^{-1}$, FWHM$\pm$10cm$^{-1}$)}\\
\hline
D band                          &                       &                               &                       &                               &                               &                               &                       &\\
Position                                &1349.0         &1358.4                 &1356.6         &1362.5                 &1362.1                 &1355.9                 &1350.0         &1342.6\\
FWHM                            &285.4          &267.0                  &236.2          &252.6                  &302.8                  &274.8                  &235.3          &279.6\\
G band                          &                       &                               &                       &                               &                               &                               &       &\\
Position                                &1595.9         &1596.5                 &1589.9         &1596.8                 &1584.8                 &1585.4                 &1586.2         &1578.8\\
FWHM                            &90.6           &89.7                   &150.4          &100.3                  &125.3                  &147.5                  &143.7          &120.5\\
\hline
\multicolumn{9}{l}{Infrared measurements}\\
\hline
$\int$C=C (cm$^{-1}$)   &17.0$\pm$7.7                   &27.1$\pm$10.9          &43.7$\pm$7.7           &13.0$\pm$4.9           &16.7$\pm$11.5          &22.7$\pm$4.3                   &12.1$\pm$5.4                           &33.0$\pm$13.6\\
$\int$C=O (cm$^{-1}$)   &2.2$\pm$0.15                   &7.5$\pm$0.5                    &29.5$\pm$0.2           &1.45$\pm$0.15  &2.3$\pm$0.15                   &8.6$\pm$0.55                   &5.3$\pm$0.35                           &9.0$\pm$0.2\\
C=O     position (cm$^{-1}$)                    &1685.7$\pm$7.0         &1693.6$\pm$7.8         &1685.1$\pm$12.5        &1687.0$\pm$5.1 &1685.1$\pm$3.8         &1700.6$\pm$12.7                &1705.9$\pm$18.5                        &1698.8$\pm$7.4\\
$\int$CH(cm$^{-1}$)             &3.62$\pm$1.21          &0.84$\pm$1.09          &0.73$\pm$0.12  &2.41$\pm$1.54  &2.2$\pm$1.9                    &1.3$\pm$0.6                    &0.66$\pm$0.43                  &0.99$\pm$2.15\\
$\int$Silicates(cm$^{-1}$)      &52.77$\pm$15.84                &35.93$\pm$10.78                &17.75$\pm$5.33 &8.33$\pm$2.50  &6.85$\pm$2.06          &3.95$\pm$1.18          &3.95$\pm$1.18                          &0.02$\pm$0.006\\
\hline
\multicolumn{9}{l}{N/C atomic ratios determined with the electron microprobe ($\pm$1$\sigma$~$^a$)}\\
\hline
N (at\%)                &-                      &3.3$\pm$1.5            &10.2$\pm$2.85  &-                      &12.1$\pm$2.5           &5.08$\pm$1.5           &-                      &11.0$\pm$2.8   \\
C (at\%)                &-                      &58.5$\pm$14.1  &76.0$\pm$5.35  &-                      &71.7$\pm$3.9           &87.2$\pm$3.3           &-                      &74.7$\pm$6.3   \\
N/C             &-                      &0.057$\pm$0.029        &0.135$\pm$0.039         &-                      &0.170$\pm$0.036        &0.058$\pm$0.017         &-                      &0.147$\pm$0.040 \\
\#measurements          &-                      &43                             &31                             &-                      &20                         &13                             &-                      &22      \\
\hline
\end{tabular}
\end{center}
$^a$ The standard deviation includes the measurements errors and intrinsic UCAMM variations of the N \& C atomic percent in the different zones probed.\\
\label{Table_summary}
\end{sidewaystable*}

\begin{thebibliography}{99}


\bibitem[Alexander et al.(2017)]{Alexander2017} Alexander, C.~M.~O.~'., Cody, G.~D., De Gregorio, B.~T., Nittler, L.~R., \& Stroud, R.~M.\ 2017, Chemie der Erde / Geochemistry, 77, 227 

\bibitem[Al{\'e}on(2010)]{Aleon2010} Al{\'e}on, J.\ 2010, \apj, 722, 1342 

%
\bibitem[Alexander et al.(2007)]{Alexander2007} Alexander, C.~M.~O.~'., Fogel, M., Yabuta, H., \& Cody, G.~D.\ 2007, \gca, 71, 4380 

\bibitem[Aug{\'e} et al.(2016)]{Auge2016} Aug{\'e}, B., Dartois, E., Engrand, C., et al.\ 2016, \aap, 592, A99 

\bibitem[Bardin et al.(2014)]{Bardin2014} Bardin, N., Slodzian, G., Wu, T.-D., et al.\ 2014, Lunar and Planetary Science Conference, 45, 2647  

\bibitem[Bergin et al.(2015)]{Bergin2015} Bergin, E.~A., Blake, G.~A., Ciesla, F., Hirschmann, M.~M., \& Li, J.\ 2015, Proceedings of the National Academy of Science, 112, 8965 

\bibitem[Bonnet et al.(2015)]{Bonnet2015} Bonnet, J.-Y., Quirico, E., Buch, A., et al.\ 2015, \icarus, 250, 53 

\bibitem[Bowey \& Adamson(2002)]{Bowey2002} Bowey, J.~E., \& Adamson, A.~J.\ 2002, \mnras, 334, 94 

%
\bibitem[Brown et al.(2011)]{Brown2011} Brown, M.~E., Burgasser, A.~J., \& Fraser, W.~C.\ 2011, \apjl, 738, L26 

\bibitem[Busemann et al.(2011)]{Busemann2011} Busemann, H., Spring, N.~H., Alexander, C.~M.~O., Nittler, L.~R.\ 2011.\ Raman Spectroscopy on Cometary and Meteoritic Organic Matter.\ Spectroscopy Letters 44, 554-559. 

\bibitem[Busemann et al.(2009)]{2009E&PSL.288...44B} Busemann, H., Nguyen, A.~N., Cody, G.~D., Hoppe, P., Kilcoyne, A.~L.~D., Stroud, R.~M., Zega, T.~J., Nittler, L.~R.\ 2009.\ Ultra-primitive interplanetary dust particles from the comet 26P/Grigg-Skjellerup dust stream collection.\ Earth and Planetary Science Letters 288, 44-57. 

\bibitem[Busemann et al.(2007)]{Busemann2007} Busemann, H., Alexander, M.~O., Nittler, L.~R.\ 2007.\ Characterization of insoluble organic matter in primitive meteorites by microRaman spectroscopy.\ Meteoritics and Planetary Science 42, 1387-1416. 

\bibitem[Charon et al.(2017)]{Charon2017} Charon, E., Engrand, C., Benzerara, K., et al.\ 2017, Lunar and Planetary Science Conference, 48, 2085 

%
\bibitem[Craig(1974)]{Craig1974} Survey of data sources on sulfide phase equilibria. In Sulfide Mineralogy (ed. P.H. Ribbe), pp. CS 3-23. Reviews in Mineralogy I , Mineralogical Society of America, Washington, D.C., USA.

\bibitem[Dartois(1998)]{Dartois1998} Dartois, E.\ 1998, Ph.D.~Thesis,  

\bibitem[Dartois et al.(2013)]{Dartois2013} Dartois, E., and 16 colleagues 2013.\ UltraCarbonaceous Antarctic micrometeorites, probing the Solar System beyond the nitrogen snow-line.\ Icarus 224, 243-252. 

\bibitem[Dartois et al.(2014)]{Dartois2014} Dartois, E., and 16 colleagues\ 2014, Geochemical Journal, 48

%
\bibitem[Dartois et al.(2015)]{Dartois2015} Dartois, E., Alata, I., Engrand, C., et al.\ 2015, Bulletin de la Societe Royale des Sciences de Liege, 84, 7 

\bibitem[Dobrica et al.(2010)]{Dobrica2010} Dobrica, E., Engrand, C., Duprat, J., \& Gounelle, M.\ 2010, Meteoritics and Planetary Science Supplement, 73, 5213 

\bibitem[Dobric{\v a} et al.(2011)]{Dobrica2011} Dobric{\v a}, E., Engrand, C., Quirico, E., Montagnac, G., Duprat, J.\ 2011.\ Raman characterization of carbonaceous matter in CONCORDIA Antarctic micrometeorites.\ Meteoritics and Planetary Science 46, 1363-1375. 

\bibitem[Draine(2015)]{Draine2015} Draine, B.~T.\ 2015, IAU General Assembly, 22, 2253136 

\bibitem[Duprat et al.(2006)]{Duprat2006} Duprat, J., Engrand, C., Maurette, M., et al.\ 2006, Meteoritics and Planetary Science Supplement, 41, 5239 

\bibitem[Duprat et al.(2007)]{Duprat2007} Duprat, J., Engrand, C., Maurette, M., et al.\ 2007, Advances in Space Research, 39, 605 

\bibitem[Duprat et al.(2010)]{Duprat2010} Duprat, J., Dobric{\u a}, E., Engrand, C., et al.\ 2010, Science, 328, 742 

\bibitem[Engrand et al.(2015)]{Engrand2015} Engrand, C., Benzerara, K., Leroux, H., et al.\ 2015, Lunar and Planetary Science Conference, 46, 1902 

\bibitem[Engrand et al.(2017)]{Engrand2017} Engrand, C., Duprat, J., Dartois, E., Godard, M., \& Delauche, L.\ 2017, EGU General Assembly Conference Abstracts, 19, 9979 

\bibitem[Fanchini et al.(2002)]{Fanchini2002} Fanchini, G., Tagliaferro, A., Conway, N.~M., \& Godet, C.\ 2002, \prb, 66, 195415 

\bibitem[Gail \& Trieloff(2017)]{Gail2017Spatial} Gail, H.-P., \& Trieloff, M.\ 2017, \aap, 606, A16 

\bibitem[Gerakines et al.(2004)]{Gerakines2004} Gerakines, P.~A., Moore, M.~H., \& Hudson, R.~L.\ 2004, \icarus, 170, 202 

\bibitem[van Ginneken et al.(2012)]{vanGinneken2012} van Ginneken, M., Folco, L., Cordier, C., \& Rochette, P.\ 2012, Meteoritics and Planetary Science, 47, 228 

\bibitem[Greshake et al.(1998)]{Greshake1998} Greshake, A., Kloeck, W., Arndt, P., et al.\ 1998, Meteoritics and Planetary Science, 33, 267

\bibitem[Halliday(2013)]{Halliday2013} Halliday, A.~N.\ 2013, Geochimica Et Cosmochimica Acta 105, 146

\bibitem[Hamilton et al.(2016)]{Hamilton2016} Hamilton, D.~P., Stern, S.~A., Moore, J.~M., \& Young, L.~A.\ 2016, \nat, 540, 97 

\bibitem[Hammer et al.(2000)]{Hammer2000} Hammer, P., Lacerda, R.~G., Droppa, R., Jr., \& Alvarez, F.\ 2000, Diamond and Related Materials, 9, 577 

\bibitem[Janches et al.(2006)]{Janches2006} Janches, D., Heinselman, C.~J., Chau, J.~L., Chandran, A., \& Woodman, R.\ 2006, Journal of Geophysical Research (Space Physics), 111, A07317 

\bibitem[Jenkins(2014)]{Jenkins2014} Jenkins, E.~B.\ 2014, arXiv:1402.4765 

\bibitem[Jenkins \& Tripp(2011)]{Jenkins2011} Jenkins, E.~B., \& Tripp, T.~M.\ 2011, \apj, 734, 65 

\bibitem[Jenkins(2009)]{Jenkins2009} Jenkins, E.~B.\ 2009, \apj, 700, 1299 

\bibitem[Jones(2015)]{Jones2015} Jones, A.\ 2015, arXiv:1511.07988 

\bibitem[Kebukawa et al.(2011)]{Kebukawa2011} Kebukawa, Y., Alexander, C.~M.~O.~'., Cody, G.~D.\ 2011.\ Compositional diversity in insoluble organic matter in type 1, 2 and 3 chondrites as detected by infrared spectroscopy.\ Geochimica et Cosmochimica Acta 75, 3530-3541. 

\bibitem[Kemper et al.(2004)]{Kemper2004} Kemper, F., Vriend, W.~J., \& Tielens, A.~G.~G.~M.\ 2004, \apj, 609, 826 

\bibitem[Kerridge(1985)]{Kerridge1985} Kerridge, J.~F.\ 1985, \gca, 49, 1707 

\bibitem[Kouketsu et al.(2014)]{Kouketsu2014} Kouketsu, Y.,\ 2014.\ Island Arc 23, 33-50.

\bibitem[Lazar et al.(2008)]{Lazar2008} Lazar, G., Bouchet-Fabre, B., Zellama, K., et al.\ 2008, Journal of Applied Physics, 104, 073534 

\bibitem[Lee et al.(2010)]{Lee2010} Lee, J.-E., Bergin, E.~A., \& Nomura, H.\ 2010, \apjl, 710, L21 

\bibitem[Love \& Brownlee(1991)]{Love1991} Love, S.~G., \& Brownlee, D.~E.\ 1991, \icarus, 89, 26 

\bibitem[Love \& Brownlee(1993)]{Love1993} Love, S.~G., \& Brownlee, D.~E.\ 1993, Science, 262, 550 

\bibitem[Marty(2012)]{Marty2012} Marty, B.\ 2012 Earth and Planetary Science Letters 313, 56

\bibitem[Matrajt et al.(2005)]{Matrajt2005} Matrajt, G., Mu{\~n}oz Caro, G.~M., Dartois, E., et al.\ 2005, \aap, 433, 979 

\bibitem[Matrajt et al.(2005)]{Matrajt2005} Matrajt, G., Brownlee, D.~E., Joswiak, D.~J., \& Taylor, S.\ 2005, 36th Annual Lunar and Planetary Science Conference, 36.

\bibitem[Millar(2015)]{Millar2015} Millar, T.~J.\ 2015, Plasma Sources Science Technology, 24, 043001 

\bibitem[Min et al.(2003)]{Min2003} Min, M., Hovenier, J.~W., \& de Koter, A.\ 2003, \aap, 404, 35 

\bibitem[Nakamura et al.(2005)]{Nakamura2005Mineralogy} Nakamura, T., Noguchi, T., Ozono, Y., Osawa, T., \& Nagao, K.\ 2005, Meteoritics and Planetary Science Supplement, 40, 5046 

\bibitem[Nesvorn{\'y} et al.(2011)]{Nesvorny2011} Nesvorn{\'y}, D., Vokrouhlick{\'y}, D., Pokorn{\'y}, P., \& Janches, D.\ 2011, \apj, 743, 37 

\bibitem[Nesvorn{\'y} et al.(2010)]{Nesvorny2010} Nesvorn{\'y}, D., Jenniskens, P., Levison, H.~F., et al.\ 2010, \apj, 713, 816 

\bibitem[Ong(1996)]{Ong1996} Ong, C.\ 1996, Thin Solid Films, 280, 1 

\bibitem[Orthous-Daunay et al.(2013)]{Orthous2013} Orthous-Daunay, F.-R., Quirico, E., Beck, P., Brissaud, O., Dartois, E., Pino, T., Schmitt, B.\ 2013.\ Mid-infrared study of the molecular structure variability of insoluble organic matter from primitive chondrites.\ Icarus 223, 534-543. 

\bibitem[Piso et al.(2016)]{Piso2016} Piso, A.-M.~A., Pegues, J., \& {\"O}berg, K.~I.\ 2016, \apj, 833, 203 

\bibitem[Poppe et al.(2011)]{Poppe2011} Poppe, A., James, D., \& Hor{\'a}nyi, M.\ 2011, \planss, 59, 319 

\bibitem[Poppe(2016)]{Poppe2016} Poppe, A.~R.\ 2016, \icarus, 264, 369 

\bibitem[Prasad et al.(2013)]{Prasad2013} Prasad, M.~S., Rudraswami, N.~G., \& Panda, D.~K.\ 2013, Journal of Geophysical Research (Planets), 118, 2381 

\bibitem[Protopapa et al.(2016)]{Protopapa2016} Protopapa, S., Grundy, W.~M., Reuter, D.~C., et al.\ 2016, arXiv:1604.08468 

\bibitem[Przybilla et al.(2013)]{Przybilla2013} Przybilla, N., Nieva, M.~F., Firnstein, M., \& Butler, K.\ 2013, EAS Publications Series, 64, 37 

\bibitem[Quirico et al.(2008)]{Quirico2008} Quirico, E., Montagnac, G., Lees, V., et al.\ 2008, \icarus, 198, 218 

\bibitem[Quirico et al.(2014)]{Quirico2014} Quirico, E., Orthous-Daunay, F.-R., Beck, P., et al.\ 2014, \gca, 136, 80 

\bibitem[Rietmeijer(2004)]{Rietmeijer2004} Rietmeijer, F.~J.~M.\ 2004, Meteoritics and Planetary Science, 39, 1869 

\bibitem[Rodil et al.(2001)]{Rodil2001} Rodil, S.~E., Ferrari, A.~C., Robertson, J., \& Milne, W.~I.\ 2001, Journal of Applied Physics, 89, 5425 

\bibitem[Rotundi et al.(2008)]{Rotundi2008} Rotundi, A., and 19 colleagues 2008.\ Combined micro-Raman, micro-infrared, and field emission scanning electron microscope analyses of comet 81P/Wild 2 particles collected by Stardust.\ Meteoritics and Planetary Science 43, 367-397. 

\bibitem[Sadezky et al.(2005)]{Sadezky2005} Sadezky, A., Muckenhuber, H., Grothe, H., Niessner, R., Poschl, U.\ 2005.\ Raman microspectroscopy of soot and related carbonaceous materials: Spectral analysis and structural information. Carbon 43, 1731-42.

\bibitem[Sandford et al.(2006)]{2006Sci...314.1720S} Sandford, S.~A., and 54 colleagues 2006.\ Organics Captured from Comet 81P/Wild 2 by the Stardust Spacecraft.\ Science 314, 1720. 

\bibitem[Suttle et al.(2017)]{Suttle2017} Suttle, M.~D., Genge, M.~J., Folco, L., \& Russell, S.~S.\ 2017, \gca, 206, 112 

\bibitem[Taylor et al.(1996)]{Taylor1996} Taylor, S., Lever, J.~H., \& Harvey, R.~P.\ 1996, Meteoritics and Planetary Science Supplement, 31,  

\bibitem[Taylor et al.(2008)]{Taylor2008} Taylor, S., Alexander, C.~M.~O., \& Wengert, S.\ 2008, Lunar and Planetary Science Conference, 39, 1628 

\bibitem[Thomas et al.(1993)]{Thomas1993} Thomas, K.~L., Blanford, G.~E., Keller, L.~P., Klock, W., \& McKay, D.~S.\ 1993, \gca, 57, 1551 

\bibitem[Toppani et al.(2001)]{Toppani2001} Toppani, A., Libourel, G., Engrand, C., \& Maurette, M.\ 2001, Meteoritics and Planetary Science, 36, 1377 

\bibitem[Verstraete(2011)]{Verstraete2011} Verstraete, L.\ 2011, European Physical Journal Web of Conferences, 18, 01001 

\bibitem[Wang et al.(2014)]{Wang2014} Wang, S., Li, A., \& Jiang, B.~W.\ 2014, \planss, 100, 32

\bibitem[Yabuta et al.(2012)]{Yabuta2012} Yabuta, H., Itoh, S., Noguchi, T., et al.\ 2012, Lunar and Planetary Science Conference, 43, 2239 

\bibitem[Yabuta et al.(2017)]{Yabuta2017Formation} Yabuta, H., Noguchi, T., Itoh, S., et al.\ 2017, \gca, 214, 172 

\bibitem[Zolensky et al.(2006)]{Zolensky2006} Zolensky, M., Bland, P., Brown, P., \& Halliday, I.\ 2006, Meteorites and the Early Solar System II, 869 

\end{thebibliography}
\end{document}